\documentclass[pra,twocolumn,preprintnumbers,amsmath,amssymb,superscriptaddress]{revtex4-2}
\usepackage{dsfont}
\usepackage{color}
\usepackage{graphicx}
\usepackage{dcolumn}
\usepackage{bm}
\usepackage{hyperref}
\usepackage{enumitem}
\usepackage{textcomp}
\usepackage{balance}
\usepackage{comment}
\usepackage{cleveref}
\usepackage{bbold}
\usepackage{amsthm}

\usepackage[commandnameprefix=ifneeded]{changes}
\usepackage[table,xcdraw]{xcolor}
\usepackage{changes}  
\renewcommand{\added}[2][]{#2}
\renewcommand{\deleted}[2][]{}
\renewcommand{\replaced}[3][]{#2}

\renewcommand{\comment}[2][]{}
\begin{document}

\title{Distillation of supersinglet states}

\author{Saeed Ahmad}

\affiliation{State Key Laboratory of Precision Spectroscopy, School of Physical and Material Sciences, East China Normal University, Shanghai 200062, China}
\affiliation{New York University Shanghai, 567 West Yangsi Road, Pudong, Shanghai 200126, China}

\author{Shuang Li} 

\affiliation{State Key Laboratory of Precision Spectroscopy, School of Physical and Material Sciences, East China Normal University, Shanghai 200062, China}
\affiliation{New York University Shanghai, 567 West Yangsi Road, Pudong, Shanghai 200126, China}

\author{Jonathan Raghoonanan} 
\affiliation{New York University Shanghai, 567 West Yangsi Road, Pudong, Shanghai 200126, China}
\affiliation{Department of Physics, New York University, New York, NY, 10003, USA}

\author{Kaixuan Zhou} 
\affiliation{New York University Shanghai, 567 West Yangsi Road, Pudong, Shanghai 200126, China}
\affiliation{Department of Physics, New York University, New York, NY, 10003, USA}

\author{Valentin Ivannikov} 
\affiliation{New York University Shanghai, 567 West Yangsi Road, Pudong, Shanghai 200126, China}

\author{Tim Byrnes}
\email{tim.byrnes@nyu.edu}
\affiliation{Department of Physics, New York University, New York, NY, 10003, USA}
\affiliation{Center for Quantum and Topological Systems (CQTS), NYUAD Research Institute, New York University Abu Dhabi, UAE}
\affiliation{New York University Shanghai, 567 West Yangsi Road, Pudong, Shanghai 200126, China}

\date{\today}

\begin{abstract}
We introduce an entanglement distillation (purification) protocol for supersinglet states composed of $ N $ qubits.  The supersinglet state we target is a total spin zero state with zero spin variance and has a fully entangled structure involving all qubits.  In our distillation protocol, three copies of an initial spin zero state are measured in the local total spin basis such that a higher fidelity supersinglet state is generated upon postselection.  The initial state can be prepared using conventional Bell state distillation methods distributed in a way to target the supersinglet symmetries.  The protocol uses only local operations and classical communication, and is suitable for long-distance applications such as quantum clock synchronization and cryptography, and avoids a high dimensional Schur transform such that it can be used for tasks such as quantum metrology. 
\end{abstract}

\maketitle

\section{Introduction}

\label{sec:intro}

Entanglement distillation is a fundamental protocol that prepares an entangled state with improved purity starting from multiple noisy copies of the same state. In the version introduced by Bennett and co-workers \cite{bennett1996} (the BBPSSW protocol), one starts with two copies of a noisy Bell state which are shared by two distant parties, Alice and Bob.  The two parties then perform a local measurement on their qubits, which projects the state to the single qubit space.  The use of measurement is an essential step in the protocol as it is an entropy reducing process, such that multiple applications of the protocol converge  towards a pure Bell state. An important constraint of entanglement distillation protocols is that they must only use local operations and classical communications (LOCC), since it is assumed that Alice and Bob are distant and entangling operations such as a CNOT gate are impossible to perform.  If such entangling operations were available, entanglement could be more easily produced by simply applying a gate.  

Since the introduction of the BBPSSW protocol, numerous generalizations of entanglement distillation/purification have been performed.  Deutsch and co-workers introduced another recurrence based protocol (the DEJMPS protocol)  with improved convergence \cite{deutsch1996quantum}.  The BBPSSW/DEJMPS recurrence protocols have been shown to be extendable to GHZ and other graph states \cite{murao1998multiparticle,maneva2000,dur2003multiparticle}. 
In Ref. \cite{kruszynska2006entanglement}, breeding and recurrence protocols were introduced to distill arbitrary graph states directly. The related task of entanglement concentration takes many partially entangled pure states to produce a few maximally entangled states \cite{bennett1996c,lo2001concentrating}.  An alternative approach for entanglement distillation/purification is based on quantum error correction used in stabilizer states which protects and transmits entangled states \cite{bennett1996mixed,matsumoto2003conversion}. Such quantum error correcting methods were extended to purify a broad spectrum of multipartite entangled states, specifically two-color graph states \cite{schlingemann2001quantum,aschauer2005multiparticle}. Generalizations to qudits were performed in Refs.  \cite{vollbrecht2003efficient,alber2001efficient}.   Since most entanglement distillation protocols are limited to purifying a particular state, a generalized and broad technique was introduced that deals with any complex multipartite stabilizer state using error correcting code \cite{glancy2006entanglement}. 
Bound entanglement introduces the notion of classes of entangled states that cannot be used for distillation \cite{horodecki1998mixed}.  
Experimentally, entanglement distillation was demonstrated via filtering \cite{kwiat2001experimental}, followed by full experimental demonstration of  entanglement distillation with photons \cite{pan2003experimental}. Other systems have also demonstrated entanglement distillation, such as solid-state systems \cite{kalb2017entanglement}. 

The class of states that allow for entanglement distillation under the LOCC restriction is rather limited, with non-stabilizer state purification protocols being more difficult to find.  A notable exception is Miyake and Briegel's W-state entanglement distillation protocol \cite{miyake2005distillation}.  Here, three noisy copies of a W-state are prepared and local measurements in a special W-basis are performed.  Generalizations of the W-state distillation protocol, to other related states, such as Dicke states, have been difficult to perform. The other class of non-stabilizer purification protocols are those for continuous variable entanglement \cite{browne2003driving,eisert2004distillation}.  Another parallel direction is magic state distillation \cite{bravyi2005universal}, which are non-stabilizer states, but generally work using error correcting methods, and the restriction on the operations are Clifford operations, rather than LOCC.  Another approach is via quantum state purification, based on the SWAP gate \cite{childs2025streaming,li2024optimal}.  While this method works on an arbitrary state, it is not an LOCC protocol; hence, it is unsuitable for distantly separated parties.  

In this paper, we present a LOCC protocol for distilling supersinglets \cite{cabello2003supersinglets}. Supersinglets are defined as the totally antisymmetric states formed by a multipartite system of qudits.  Specifically, we refer to supersinglet states that are formed in $ N $ qubit systems (assumed to be an even number throughout this paper).  For two qubits, there is only one singlet state: 
\begin{align}
   |\Psi^- \rangle =  (|01\rangle - |10 \rangle)/\sqrt{2} . 
\end{align} 
Successively coupling the spin-1/2 qubits together, for more than two qubits $N \ge 4 $, there are more than one multiplicity of singlet state.  One may always form a spin zero state for any even $ N $ by taking a simple product state of singlet states $ |\Psi^- \rangle^{\otimes N/2} $. These are however not fully entangled states and do not possess quantum correlations that are useful for applications such as quantum clock synchronization or cryptography. 
A supersinglet may be formed by coupling half of the qubits form a maximal spin $ s = N/4 $, then coupling these antiferromagnetically to produce a total spin zero state.  In the supersinglet language, half the qubits form an effective $ N/2+1 $ dimensional qudit,  and there are two such qudits.  This produces a fully entangled state in the sense that all the qubits participate in the entangled state.

Purifying supersinglet states is non-trivial because they are non-stabilizer states for $ N \ge 4 $. They generally have a non-uniform superposition, which makes a generalization of BBPSSW/DEJMPS methods more difficult to apply.  Quantum error correction methods are also difficult to apply due to the fact that they are non-stabilizer states, and is challenging to find a code where the supersinglet is in the codespace for all $ N $.  They are however potentially useful in quantum information applications since the variance of the total spin in any basis is also zero, hence they are an example of a state with zero quantum noise.  Applications of supersinglets include cryptography \cite{cabello2003supersinglets}, quantum clock synchronization \cite{jozsa2000,ilo-okeke2018}, quantum metrology \cite{toth2010generation,chaudhury2007quantum,PhysRevA.106.033314}, quantum teleportation \cite{horodecki1999general,bennett1996purification,pyrkov2014full}, quantum computing \cite{byrnes2012,abdelrahman2014coherent}, and decoherence free subspaces \cite{toth2010generation}. 



\section{Physical system}

\subsection{Qubit configuration}

The physical system that we will consider is shown in Fig. \ref{fig1}. There are a total of $3N $ qubits, corresponding to three copies of the quantum state $ \rho $, consisting of $ N $ qubits each.  We will take $ N $ to be even throughout this paper, as we aim to produce supersinglets, which only exist for $N $ even. 
In Fig. \ref{fig1}, each of the rows corresponds  to one copy of the quantum state which eventually will be distilled to a 
supersinglet.  Each local party (Alice, Bob, Charlie, \dots) is labeled by an index $ n \in [1,N] $. The labels $ d \in [1,3] $ identify the copy of the supersinglet. Each of the parties $ n $ may be separated by a large distance, while the three qubits for each duplicate state are assumed to be at the same location.  For this reason we call qubits with the same $ n $ to be ``local'', such that there is no restriction on the operations that can be performed. Qubits with different $ n $ are restricted to LOCC operations, since they are considered to be distant.    

The Pauli operators for the qubit labeled by $ (n,d) $ are 
\begin{align}
\vec{\sigma}_{nd} = (\sigma^x_{nd}, \sigma^y_{nd},\sigma^z_{nd})  .
\end{align}
We now define the relevant basis states that will be used in our distillation protocol.

\subsection{Basis within each duplicate}

\begin{figure}[t!]
\includegraphics[width=\linewidth]{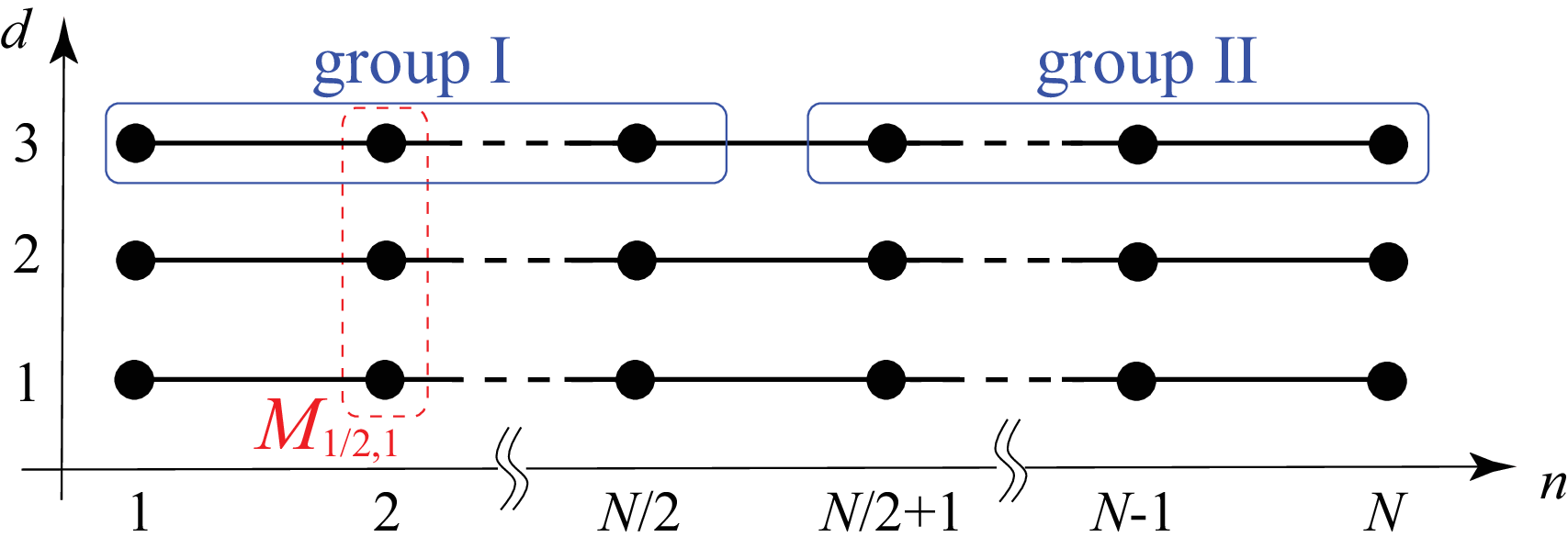}
	\caption{Qubit configuration for our distillation protocol for supersinglets. 
    Each row of qubits contains contain $ N $ qubits (assumed even), shared between distant parties labeled by $ n $.  At the end of the supersinglet distillation protocol, each row stores a supersinglet state $ | {\cal S}_N \rangle $.      
    Each party holds 3 qubits, forming three duplicates of the quantum state. A local measurement corresponding to the measurement operator (\ref{measurementandrot}) is performed at each local site (dashed box) and postselected to the result $ j= 1/2, \alpha = 1 $.  Horizontal boxes label group I and II qubits which define the symmetry of the supersinglet state defined by (\ref{perm1}) and (\ref{perm2}).    }
	\label{fig1}%
\end{figure}

Here, we define the basis within each duplicate state in our array of qubits.  This corresponds to each of the rows of Fig. \ref{fig1}, labeled by $ d $. After purification, each of these rows will store the supersinglet state. 
In each row, there are $ N$ qubits, for which we define the total spin operators
\begin{align}
\vec{S} = \frac{1}{2} \sum_{n=1}^{N} \vec{\sigma}_{nd} . 
\end{align}
For notational simplicity, we will suppress the duplication label $ d $ on the operator $ \vec{S} $, and implicitly work with a fixed $ d $.  The total angular momentum eigenstates $ | s,\alpha,m \rangle $ satisfy the eigenvalue equations
\begin{align}
S^2 | s,\alpha ,m \rangle & = s(s+1) | s,\alpha,m \rangle  \nonumber \\
S^z | s,\alpha,m \rangle & =  m | s,\alpha,m \rangle 
\label{spinseigenstates}
\end{align}
where $ \alpha $ is the outer multiplicity label and $ S^2 = \vec{S} \cdot \vec{S} $.  For even $ N $, the spins take values $ s \in \{0,1,\dots,N/2 \} $.  
The $S^z $ eigenvalue has a range $ m \in [-s,s] $ and the multiplicity label has a range $ \alpha \in [1,A(N,s)] $, where 

\begin{align}
A(N,s) = \binom{N}{N/2-s} - \binom{N}{N/2-s-1} .
\end{align}
In particular, the singlet sector $ s = 0 $ has a multiplicity of
\begin{align}
    A(N,0) = \frac{\binom{N}{N/2}}{N/2+1} .
    \label{singletmult}
\end{align}

\subsection{Local basis}

Here we define the local basis at each site located at $ n $. This corresponds to the columns of Fig. \ref{fig1}. 
We define the total spin for the three qubits held by each party $ n $ as
\begin{align}
\vec{J} = \frac{1}{2} \sum_{d=1}^{3} \vec{\sigma}_{nd} .
\end{align}
Again, for notational simplicity, we will suppress the site label $ n $ on the operator $ \vec{J} $, and implicitly work with a particular local site $ n $.

The angular momentum eigenstates are defined in the same way as (\ref{spinseigenstates}), 
\begin{align}
J^2  | j,\alpha,m \rangle & = j(j+1) | j,\alpha,m \rangle  \nonumber \\
J^z  | j,\alpha,m \rangle & =  m | j,\alpha,m \rangle .
\label{spinseigenstatesj}
\end{align}

From standard angular momentum coupling rules, three spin-1/2 (qubits) couple to form three irreducible representations $ \tfrac{1}{2} \otimes \tfrac{1}{2} \otimes \tfrac{1}{2} \cong \tfrac{3}{2} \oplus  \tfrac{1}{2} \oplus  \tfrac{1}{2} $. Of these sectors, we will be particularly interested in the spin sector $ j = 1/2, \alpha = 1 $:
\begin{align}
|0^{(3)}  \rangle& := | \tfrac{1}{2},1, \tfrac{1}{2} \rangle= \frac{1}{\sqrt{6}} (-2 |001 \rangle + |010\rangle + |100 \rangle  )  \nonumber \\
|1^{(3)} \rangle & :=  | \tfrac{1}{2},1, -\tfrac{1}{2} \rangle = \frac{1}{\sqrt{6}} ( - 2|110 \rangle +  | 101\rangle + |011 \rangle ) .
\label{3qubitlogical}
\end{align}
These states transform in the same way as a spin-1/2 under rotations of $ \vec{J} $.  The superscript $ ^{(3)} $ reminds us that the states consist of three qubits.  The full set of angular momentum eigenstates in the 3 qubit space is listed in Appendix \ref{app:3qubit}.

\section{Supersinglets}

\subsection{Definition}

We first define the supersinglet state. In the most general definition, supersinglets are completely antisymmetric states consisting of several qudits \cite{cabello2003supersinglets}.  In this paper, we only consider supersinglets that are formed from $ N $ qubits. 
First, supersinglets are a type of singlet state with total spin $ s= 0 $.
Therefore, they have eigenvalues $ s = m = 0 $ in Eq. (\ref{spinseigenstates}).  The multiplicity of such spin zero states is $ A(N,0) $, hence for $ N \ge 4 $, additional symmetries are required to specify the supersinglet state. The multiplicity of spin zero states is determined by the way that the angular momentum of the $ N $ qubits couple (see also Appendix \ref{app:3qubit}).  The supersinglet is the singlet state that couples in such a way that there are two groups (called group I and II), each of $ N/2 $ qubits, such that each group forms the  maximal spin state of $ s = N/4 $:
\begin{align}
S_{\text{I}}^2 | {\cal S} \rangle = S_{\text{II}}^2 |  {\cal S} \rangle = \frac{N}{4} ( \frac{N}{4} + 1 ) |  {\cal S}\rangle .
\label{maximalspin}
\end{align}
Here, spin operators for each half of the qubits are defined as
\begin{align}
\vec{S}_{\text{I}} & = \frac{1}{2}  \sum_{n=1}^{N/2} \vec{\sigma}_n  \nonumber \\
\vec{S}_{\text{II}} & = \frac{1}{2}  \sum_{n=N/2+1}^{N} \vec{\sigma}_n  .
\label{s1s2def}
\end{align}
The site labels of group I ($ n \in [1,N/2] $) and group II ($ n \in [N/2+1,N] $) are of course arbitrary and could be chosen with a different convention.  We will use this convention throughout this paper. 

The supersinglet state is then formed by coupling the group I spins (forming a $s=N/4$ spin) antisymmetrically with the group II spins (also forming a $s=N/4$ spin). There is only one such state due to the uniqueness of the supersinglet state \cite{cabello2003supersinglets}.  The explicit wavefunction of the supersinglet is written \cite{cabello2003supersinglets}
\begin{align}
| {\cal S}_N \rangle = \frac{1}{\sqrt{N/2+1}} \sum_{k=0}^{N/2} (-1)^k |D^{(N/2)}_k \rangle |D^{(N/2)}_{N/2-k} \rangle
\label{supersingletwave}
\end{align}
where the Dicke states can be defined as
\begin{align}
|D^{(N/2)}_k \rangle & = | \tfrac{N}{4},1, \tfrac{N}{4}-k \rangle \nonumber \\
& = \frac{1}{\sqrt{\binom{N/2}{k}}} \sum_{\sigma} P_{\sigma} \left( | 0 \rangle^{\otimes N/2-k} | 1 \rangle^{\otimes k}  \right) .
\label{dickestate}
\end{align}
The permutation operator is defined in the computational basis
%
\begin{align}
P_{\sigma} | k_1 k_2 \dots k_N \rangle = | k_{\sigma(1)} k_{\sigma(2)} \dots k_{\sigma(N)} \rangle .
\end{align}
Here, the function $ \sigma(n) $ specifies the permutation, and $ k_n \in \{0,1 \} $ for $ n \in [1,N] $.  The sum in (\ref{dickestate}) is over all possible distinct permutations of the qubits.  The Dicke state is thus a completely symmetric state under qubit interchange. \deleted{In}

We provide an explicit example for the $ N =4 $ case.  Coupling the qubits one by one, we have
\begin{align}
    1/2 \otimes 1/2 \otimes 1/2 \otimes 1/2 & = (0 \oplus 1) \otimes 1/2 \otimes 1/2 \nonumber \\
    & = (1/2 \oplus 1/2 \oplus 3/2) \otimes 1/2  \nonumber \\
 & = 0 \oplus 1 \oplus 0 \oplus 1  \oplus 1  \oplus 2 .
\end{align}
In the second line there are two $ s =1/2 $ because it can originate from coupling $ s=0 $ or $ s= 1 $ with $ s=1/2 $.  Similarly, there are two $ s=0 $ sectors in the last line from coupling the last $ s = 1/2 $ spin with the two multiplicities of $ s = 1/2 $.  The $ s = 0 $ multiplicity is therefore  $ A(N,0)=2 $, and there are two singlet states.  The supersinglet is the singlet state that satisfies (\ref{maximalspin}).  The explicit wavefunction can be written using (\ref{supersingletwave}) and is given by 
\begin{align}
    | {\cal S}_4 \rangle = & \frac{1}{2\sqrt{3}} ( -2 |0011 \rangle + |0101\rangle + |0110\rangle \nonumber \\
& + |1001\rangle + |1010\rangle  -2 | 1100 \rangle ). 
\label{supersinglet4}
\end{align}

\subsection{Basic properties}

\subsubsection{Rotational invariance}

Supersinglets, as with any $s = 0 $ state, are invariant under any spin rotation
\begin{align}
e^{-i \vec{S} \cdot \vec{w} \theta} | s=0, \alpha, m= 0 \rangle & =
U^{\otimes N} | s=0, \alpha, m= 0 \rangle  \nonumber \\
& =  | s=0, \alpha, m= 0 \rangle ,
\label{spinrot}
\end{align}
where $ U = e^{-i \vec{\sigma} \cdot \vec{w} \theta/2} $ is a single qubit rotation.  This property will be important in the context of twirling operations introduced later (Sec. \ref{sec:twirling}).

\subsubsection{Permutation symmetry}

The spatial symmetries of the supersinglet state are evident from the form of the wavefunction (\ref{supersingletwave}).  First, since Dicke states are symmetric superpositions of qubits, permutations that keep qubits within groups I and II as given in (\ref{s1s2def}) leave the supersinglet invariant:
\begin{align}
P_{\sigma_{\text{I},\text{II}}} | {\cal S}_N \rangle & = | {\cal S}_N \rangle  \label{perm1}
\end{align}
where $\sigma_{\text{I},\text{II}}$ denotes a permutation that keeps qubits within groups $ \text{I},\text{II} $ respectively (e.g. $ \sigma_{\text{I},\text{II}} = 1234,1243,2134,2143 $ in the one-line notation for $ N = 4$).  We may also interchange groups $ \text{I} $ and $ \text{II} $ such that 
\begin{align}
P_{\sigma_{\text{I} \leftrightarrow \text{II} } } | {\cal S}_N \rangle  = (-1)^{N/2} | {\cal S}_N \rangle ,
\label{perm2}
\end{align}
which is the same state up to an irrelevant global phase.  Here $ \sigma(n) = (n-1+ N/2 (\text{mod} N)) + 1 $ (e.g. $ \sigma_{\text{I} \leftrightarrow \text{II} } = 3412 $ for $ N = 4 $).

\subsubsection{Spin variances}

As spin zero states, supersinglets have zero variance in all directions
\begin{align}
\text{Var} (S^i) = \langle {\cal S}_N | (S^i)^2 | {\cal S}_N \rangle - 
\langle {\cal S}_N | S^i | {\cal S}_N \rangle^2  = 0
\end{align}
for $ i = x,y,z $. This follows from the fact that a supersinglet is a $ s = 0 $ state and the rotational invariance (\ref{spinrot}).

\section{Allowed operations}
\label{sec:allowed}
In the preceding sections, we presented the basic properties of the physical system and supersinglets.
We now introduce the allowed operations that are used in our distillation protocol. Our primary constraint is that all operations in the main distillation sequence should only use LOCC.  

\subsection{Local measurements}

Measurements in the local basis (\ref{spinseigenstatesj}) are made which distinguish the spin and outer multiplicity label $ j, \alpha $.  Projective measurements in the three sectors $ (j = 3/2, \alpha =1)$, $ (j = 1/2, \alpha =1)$, $ (j = 1/2, \alpha =2)$, are defined as 
\begin{align}
\Pi_{j \alpha } = \sum_{m=-j}^j | j,\alpha , m \rangle \langle j, \alpha , m | .
\label{projections}
\end{align}
Since these are projectors, they satisfy
\begin{align}
\sum_{j \alpha }  \Pi_{j  \alpha } = I  ,
\end{align}
in each local space labeled by $ n $.

\subsection{Schur transform}

In the local 3 qubit basis, we perform a Schur transform \cite{bacon2006efficient}, which rotates from the total angular momentum basis to the computational basis.  For the three qubits held by each party, we define this as
\begin{align}
U_{\text{Sch}}^{(N=3)} = \sum_{j=1/2}^{3/2} \sum_{\alpha=1}^{A(3,j)} \sum_{m=-j}^j | v(j,\alpha,m) \rangle \langle  j, \alpha, m |
\label{schurunitary}
\end{align}
where $ v(j,\alpha,m) $ denotes a computational basis state labeled by the index function $v(j,\alpha,m)$. The explicit ordering convention is given in Appendix \ref{vjamordering}.












\subsection{Measurement operator} 

In our distillation protocol, we will perform the measurement (\ref{projections}), followed by the Schur transform (\ref{schurunitary}).  We may define the combined operation of these as the measurement operator
\begin{align}
M_{j \alpha} & = U_{\text{Sch}}^{(N=3)} \Pi_{j \alpha }  \nonumber \\
& = \sum_{m=-j}^j |  v(j,\alpha,m) \rangle \langle j, \alpha , m | .
\label{measurementandrot}
\end{align}
As we show later, we will postselect on the measurement outcome  $ j = 1/2, \alpha = 1 $.  Using our convention of the Schur transform (\ref{schurunitary}), the final state will be in the space spanned by the states $ \{ | 000 \rangle, | 001 \rangle \} $. As the first two qubits are in the fixed state $ |00 \rangle $, they may be discarded. We may define the effective measurement operator for the successful outcome

\begin{align}
M_{1/2,1} & = | 0 \rangle \langle \tfrac{1}{2}, 1,  \tfrac{1}{2} | 
+  | 1 \rangle \langle \tfrac{1}{2}, 1,  -\tfrac{1}{2} |   \nonumber \\
& = | 0 \rangle \langle 0^{(3)} |+  | 1 \rangle \langle1^{(3)} | .
\label{measureps}
\end{align}
This maps a state from the three qubit space to the single qubit space.  This operator captures the effect of one round of purification.

\subsection{Twirling}
\label{sec:twirling}

For each duplicate system (i.e. the rows of Fig. \ref{fig1}), we assume that ``twirling'' operations are possible, defined as
\begin{align}
\rho \rightarrow T(\rho) = \int dU U^{\otimes N} \rho {U^\dagger}^{\otimes N} ,
\label{twirldef}
\end{align}
where $ U = e^{-i \vec{\sigma} \cdot \vec{w} \theta /2 } $ is a single qubit rotation,  and $ \vec{w} $ is a normalized vector specifying the axis of rotation, and $ \theta $ is the rotation angle.  The $ U^{\otimes N}  $ is the generalization of the bilateral rotations that are present in protocols such as BBPSSW \cite{bennett1996}. 

We evaluate that in the total spin basis (see Appendix \ref{app:twirl}) the twirling operation corresponds to
\begin{align}
T(\rho) = \sum_s \frac{1}{2s+1} \sum_{l, l'} \text{Tr} (\rho \Gamma_{s l l'}^\dagger ) \Gamma_{s l l'}
\label{twirlformula}
\end{align}
where
\begin{align}
\Gamma_{s l l'} = \sum_{m=-s}^s |s, l, m \rangle \langle s, l', m| .
\label{gammadef}
\end{align}
The basic effect of the twirling operation can be seen to produce an even mixture of $ m $ states within each $ s, l $ sector, weighted by the original probability in that sector.  For example, in the familiar two-qubit ($N=2$) case, an arbitrary two qubit state $ \rho $ is converted to 
\begin{align}
T(\rho) =  F | 0,1,0\rangle \langle 0,1,0| + \frac{1-F}{3} \sum_{m=-1}^1 | 1,1,m\rangle \langle 1,1,m|
\end{align} 
where $ F = \langle 0, 1, 0 | \rho | 0,1,0 \rangle $ is the fidelity with respect to the singlet state. 

An important counterexample to this general rule is that the coherence between different outer multiplicity labels is not removed by twirling.  For example, consider the four qubit case ($ N = 4 $), and suppose one were to start with the state
\begin{align}
|\xi\rangle = a | 1,1,1\rangle +b |1,2,1\rangle ,
\end{align}
with complex coefficients $ a,b $.  This state has $ s = 1$, $ m = 1 $ for both but different outer multiplicity labels $ \alpha  $ in the superposition.  In this case the twirling operation acts as
\begin{align}
& T(|\xi\rangle \langle \xi |) = \nonumber \\
& \frac{1}{3} \sum_{m=-1}^1 (a | 1,1,m\rangle + b |1,2,m\rangle ) (a^* \langle 1,1,m | + b^* \langle 1,2,m |) .
\label{twirlingexample}
\end{align}
Thus while a mixture of different $ m $ are produced, the superposition between different $  \alpha $ is not erased.  This can be problematic for distillation protocols as off-diagonal terms in the density matrix tend to prevent convergence towards perfect fidelities \cite{bennett1996,deutsch1996quantum}.

\subsection{Permutations}

The last type of operation that we use is permutation symmetrization.  To target states with particular symmetries, we may apply the permutation symmetrizer
\begin{align}
\rho \rightarrow W(\rho) = \frac{1}{|Q|} \sum_{\sigma \in Q} P_{\sigma} \rho P_{\sigma}^\dagger ,
\label{permsymm}
\end{align}
where $ Q  $ is the set of permutations with the symmetry that one would like to enforce.  In the case of supersinglets, the type of symmetry that would be enforced would be (\ref{perm1}) and (\ref{perm2}).  For example, for $N =4 $ the supersinglet symmetry would be $ Q = \{1234, 1243, 2134,2143,3412,3421,4312,4321\} $.

We note that the permutation symmetrizer is not strictly local as it involves the physical interchange of qubits, or applying a sequence of SWAP gates.  Hence it is not strictly an LOCC operation, although it is incapable of producing entanglement.  For this reason it should not be used in the purification protocol itself, although it may be used in the initialization step during the distribution of undistilled states.

\section{Supersinglet distillation}

\subsection{Protocol}
\label{sec:protocol}
After introducing the allowed operations, we are now ready to present our supersinglet distillation protocol. The procedure proceeds as follows:
\begin{enumerate}
\item Prepare an initial state $ \rho$ with total spin zero, with the dominant fidelity $ F_{\alpha} =  \langle 0, \alpha, 0 | \rho | 0, \alpha, 0 \rangle $ being the target supersinglet.  This can be performed using conventional singlet Bell pair distillation, starting from noisy Bell pairs, for example.  
\item Perform the twirling operation (\ref{twirlformula}) to remove coherences between spin sectors. 
\item Perform permutation symmetrization (\ref{permsymm}) such that the state has the same symmetry as a supersinglet.
\item Take three copies of the state and perform local \added{spin} measurements (\ref{projections}) and a Schur transform (\ref{schurunitary}) at each local site $ n $. 
\item All parties classically broadcast their measurement outcomes and postselect on the outcome $ j = 1/2, \alpha = 1 $. The fixed qubits $ d=1,2 $ are discarded such that the effective operation is (\ref{measureps}).  
\item Using only the postselected state, recursively iterate Steps 4 and 5 until a high fidelity supersinglet is obtained. 
\end{enumerate}

Steps 1-3 correspond to preparing the initial state prior to the recurrence steps of the purification, which are performed in Steps 4-6.  We note that one point of difference to standard distillation protocols is that the initial state that is prepared is not a noisy supersinglet state, but a state in the spin zero sector with dominant fidelity as the supersinglet.  

Our protocol is LOCC compliant. The first step, which uses standard Bell pair distillation, only involve LOCC, except for the initial distribution of Bell states.  Steps 2 and 4-6  explicitly only use LOCC.  Step 3 uses permutations which are not local operations.  This, however, can also be considered to be part of the initial state distribution. For example, distributing singlet Bell pairs in a suitable geometry ensures that the symmetry of the supersinglet is enforced.

\subsection{Analysis of the protocol}
\label{sec:Analysis}

We now further elaborate on each step of the protocol and explicitly show the states at each step where possible.

\subsubsection{Step 1: State initialization}
\label{sec:step1init}

In the first step, we need to prepare a state with spin zero that has its dominant fidelity in the target supersinglet state.  One simple way that this can be achieved is by using conventional Bell state purification.  Starting from noisy Bell pairs, using protocols such as BBPSSW/DEJMPS  \cite{bennett1996,deutsch1996quantum} (or similar), one can prepare high fidelity singlet states.  Since a product of singlet Bell states $ |\Psi^- \rangle^{\otimes N/2} $ will always have total spin zero, this satisfies the spin zero requirement. 

To ensure that the state has a non-zero fidelity with the target supersinglet, we arrange the singlets such that one qubit is in group I and the other in group II. 
Arrangements of singlets where both qubits are in group I or II give zero overlap with the supersinglet due to the antisymmetry of singlet states.  
The state at this point is 
\begin{align}
|\psi_1 \rangle & = |\Psi^- \rangle_{1,N/2+1}  |\Psi^- \rangle_{2,N/2+2} \dots
|\Psi^- \rangle_{N/2, N}  
\label{step1state}
\end{align}
where the subscripts denote the qubit numbers $ n $ for each Bell pair. All the entanglement is {\it between} groups I and II.

The type of state that is created is a total spin zero state and satisfies
\begin{align}
S^2 |\psi_1 \rangle = 0 .
\end{align}
and 
\begin{align}
S^z |\psi_1 \rangle = 0 .
\end{align}

It is however not a supersinglet state (\ref{supersingletwave}), as it does not have the symmetries (\ref{perm1}) and (\ref{perm2}).  The state $ |\psi_1 \rangle  $  is non-orthogonal to the supersinglet state, which is important as an initial state for distillation.  

We examine some alternatives to the singlet Bell state generation scenario in Sec. \ref{sec:alternativeinitial}.

\subsubsection{Step 2: Perform twirling operations}

The next step is to perform twirling operations (\ref{twirlformula}) by averaging over local unitary rotations.  For the state (\ref{step1state}), the state is left unchanged since it is already a spin zero state:
\begin{align}
\rho_2 = T(|\psi_1 \rangle \langle \psi_1 |) = |\psi_1 \rangle \langle \psi_1 |.
\end{align}
Therefore, in case the initial state is a singlet, twirling is redundant.  However, for other choices of initial state (such as the modified GHZ state introduced later), coherences between different spin sectors are removed, assisting the convergence of the distillation algorithm.

\subsubsection{Step 3: Enforce permutation symmetry}
\label{sec:enforce}

In this step, we enforce the supersinglet symmetries (\ref{perm1}) and  (\ref{perm2}) on the initial state using the permutation symmetrizer (\ref{permsymm}).  For the state (\ref{step1state}), in fact, only (\ref{perm1}) is necessary because interchange of groups I and II \added{leaves} \deleted{leave} the state invariant up to a global phase.   

We can make another simplification by only applying permutations $ \sigma $ in group II and leaving group I unchanged
\begin{align}
\sigma= 12 \dots (N/2) \sigma_{\text{II}} 
\label{simpleperm}
\end{align}
in Eq. (\ref{permsymm}), where $ \sigma_{\text{II}} $ is any permutation of group II labels $ N/2+1 \dots N $.   The reason we can make this simplification is that the original state (\ref{step1state}) is a product of identical singlet states, hence interchanging both group I and group II qubits can result in an identical state.  For example, for $ N = 4 $, the \added{permutation symmetrized Bell (PSB)} initial state will be 
\begin{align}
\rho_{\text{PSB}} =  & \frac{1}{2} \Big(  |\Psi^- \rangle\langle \Psi^- |_{1,3}  \otimes   |\Psi^- \rangle\langle \Psi^- |_{2,4} \nonumber \\
& + |\Psi^- \rangle\langle \Psi^- |_{1,4}  \otimes   |\Psi^- \rangle\langle \Psi^- |_{2,3} \Big) .
\label{rho3state}
\end{align}
\replaced{Such a state may be produced by randomly choosing between singlets with pairing between sites 13 and 24, and singlets with pairing 14 and 23, with a probability of 1/2 for each case. }{
Clearly, such a state may be produced by initially distributing half the singlets between sites 13 and 24, and the other half with the pairing 14 and 23.  } \added{More generally, to produce the permutation symmetric mixed state, one of the permutations (\ref{simpleperm}) is chosen at random with equal probability.  For each randomly chosen configuration, the singlets are distributed according to the chosen configuration. }  In this way, an explicit SWAP operation is not necessary, \added{as the Bell state entanglement is produced in the distribution stage.} \deleted{correct symmetry can be enforced \replaced{at}{in} the entanglement distribution stage.}

We note that the initial state (\ref{rho3state}) is no longer a pure state, due to the symmetrization.  It is nevertheless a total spin zero state since all terms in the mixture have spin zero:
\begin{align}
    \text{Tr} (\rho_{\text{PSB}} S^2) =  \text{Tr} (\rho_{\text{PSB}} S^z) = 0 .
\end{align}

\subsubsection{Steps 4-5: Local measurements and postselection}

Each of the local parties performs a projective measurement in the total spin basis, followed by the Schur transform, corresponding to the measurement operator
\begin{align}
 M_{\vec{j} \vec{\alpha} } =   \bigotimes_{n=1}^N M_{j_n \alpha_n}
\end{align}
where $ j_n, \alpha_n $ are the outcomes at the $ n$th site, and $ M_{j \alpha} $ is given by (\ref{measurementandrot}).  Postselecting on the $ j = 1/2, \alpha = 1 $ outcome for each party corresponds to applying the operator
\begin{align}
{\cal M}  = M_{1/2,1}^{\otimes N}, 
\label{psoperatorbig}
\end{align}
where $M_{1/2,1}$ is given by (\ref{measureps}).   This operator takes a state with $ 3N $ qubits and outputs a state in the $ N $ qubit space.  The update procedure for an input density matrix $\rho $ is then 
\begin{align}
\rho \rightarrow \rho' = \frac{1}{p_{\text{suc}}} {\cal M} \rho^{\otimes 3} {\cal M}^\dagger
\label{rhoupdate}
\end{align}
where the probability of obtaining the desired $ j = 1/2, \alpha = 1 $ outcome is
\begin{align}
p_{\text{suc}} = \text{Tr} ( {\cal M} \rho^{\otimes 3} {\cal M}^\dagger) .
\label{generalsuccessprob}
\end{align}

The key property of the operator (\ref{psoperatorbig}) is that it leaves the supersinglet state invariant
\begin{align}
   {\cal M} | {\cal S}_N \rangle^{\otimes 3} = \sqrt{p_{\text{suc}}} | {\cal S}_N \rangle
   \label{fixedpoint}
\end{align}
where $ p_{\text{suc}} =  \langle {\cal S}_N |^{\otimes 3}  {\cal M}^\dagger  {\cal M} | {\cal S}_N \rangle^{\otimes 3}  $ in this case.  This leaves the supersinglet a fixed point of the recurrence procedure.  The argument to show that (\ref{fixedpoint}) is true for any $ N $ is shown in Appendix \ref{app:fixedpointproof}.  The basic argument relies on the isomorphism of the $ j = 1/2, \alpha = 1 $ space in the 3 qubit manifold with the 1 qubit $ j = 1/2 $ manifold. Then, because $ | {\cal S}_N \rangle^{\otimes 3} $ is itself a spin zero state with the same permutation symmetry as $ | {\cal S}_N \rangle $, and after applying Schur's lemma, they must be the same vector up to a constant.

\begin{figure}[t!]
\includegraphics[width=\linewidth]{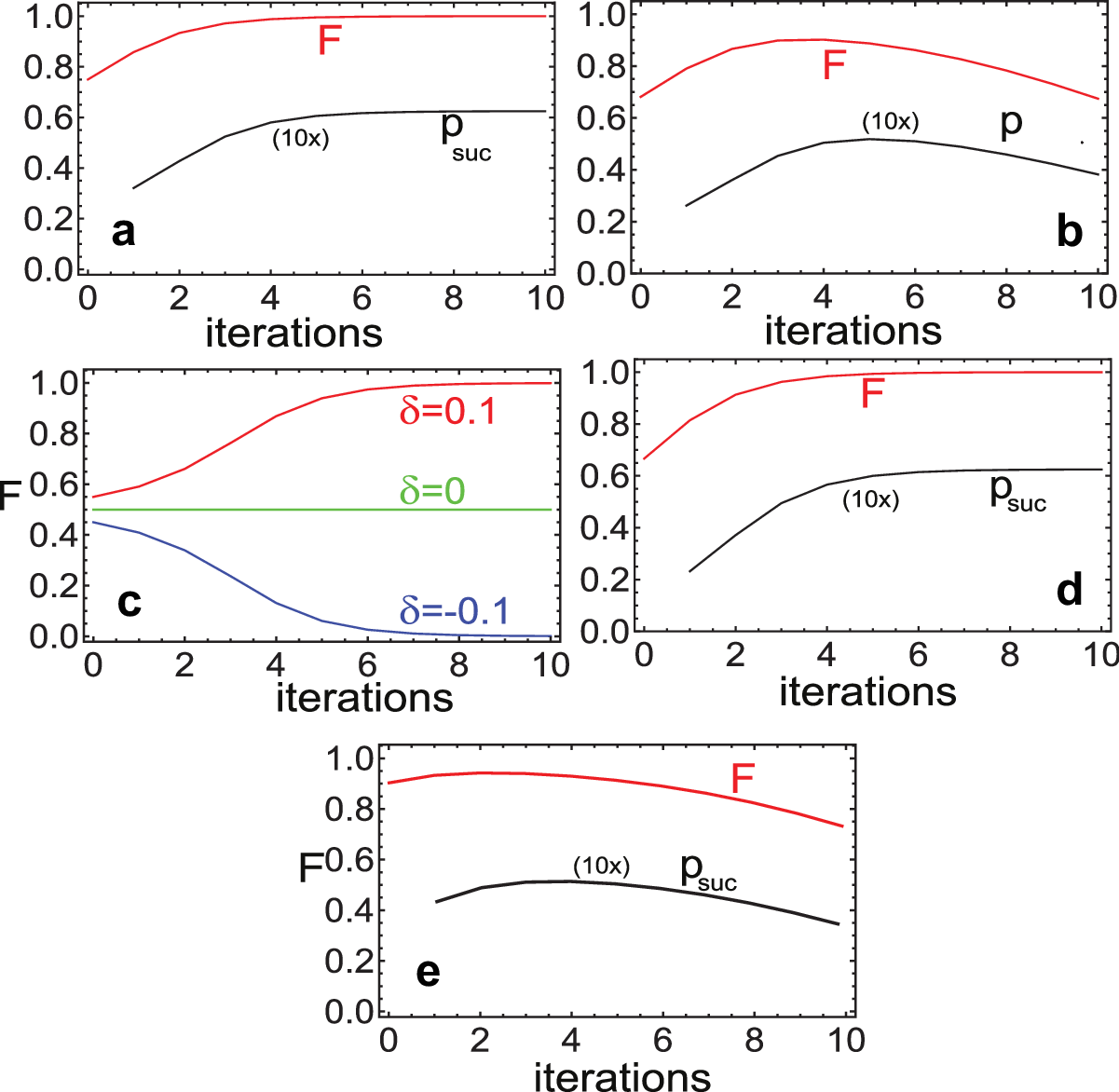}
	\caption{Numerical simulation of \replaced{the}{our} supersinglet distillation protocol \added{of Sec. \ref{sec:protocol} for $ N =4 $}.    
\deleted{Each plot shows the fidelity (\ref{fidelity}) and success probability (\ref{generalsuccessprob}) multiplied by a factor of 10 for visibility. 
    Results shown are (a) $ N = 4 $ with the initial state (\ref{rho3state});  (b) $ N = 4 $ with the initial state (\ref{wernerstate}) and $ \epsilon = 0.1 $;
    (c) fidelities for $ N = 4 $ with the initial state (\ref{s0mixture}) for $ \delta $ as marked; (d) $ N = 4 $ with the initial state (\ref{ghzlikestate}) with twirling operations (\ref{twirlformula}).  } 
    \added{For (a)(b)(d)(e) we plot the fidelity (\ref{fidelity}) and success probability (\ref{generalsuccessprob}) multiplied by a factor of 10 for visibility. The protocol given in Steps 4-6 in Sec. \ref{sec:protocol} are performed with the initial state choices of (a) the permutation symmetric Bell (PSB) state (\ref{rho3state});  (b) Werner-PSB state (\ref{wernerstate}) with $ \epsilon = 0.1 $;
    (c) $s=0 $ mixture state (\ref{s0mixture}) for $ \delta $ as marked; (d) GHZ state (\ref{ghzlikestate}) with twirling operations (\ref{twirlformula}); (e) Werner-supersinglet state (\ref{supersinglet}).   }
    }
	\label{fig2}%
\end{figure}

\section{Numerical simulation}

\replaced{In this section, we}{ We now} directly simulate the steps of the protocol as given in Sec. \ref{sec:protocol} to verify the performance of the distillation protocol. \added{The main aspect that we will investigate are the choices of initial state that will be used in the protocol, i.e. various choices in Step 1.  Specifically, we investigate effects of noise on the initial Bell state prepared in Sec. \ref{sec:step1init}-\ref{sec:enforce}, and other initial states such as Werner-supersinglets, GHZ states, as well as mixtures of $ s = 0 $ states.  
}

\subsection{\added{Methods}}

The step in the protocol that requires the largest computational overhead is (\ref{rhoupdate}), since it involves evaluation of three copies of the density matrix which have dimension $ 2^{3N} \times 2^{3N} $.  Even for $ N = 6 $ the memory requirements to store the density matrix becomes challenging on a standard computer.  

To evaluate larger systems than $ N = 4 $, we utilize the fact that for an initial state that is perfectly in the $ s = 0 $ sector, the density matrix never leaves the total spin zero sector under the distillation sequence of Sec. \ref{sec:protocol}. The reason for this is that the projector (\ref{projections}) with $ j = 1/2, \alpha = 1 $ for all $ n $ commutes with the total spin operator.  The twirling and permutation operations likewise do not change the spin sectors.  Thus if the initial state is purely in the $ s = 0 $ sector, the state will remain in the same sector for all iterations, and remaining spin sectors may be safely truncated.  In Appendix \ref{app:simulation} we show the evaluation of Eq. (\ref{rhoupdate}) with such a truncation.  For an initial state that is exactly a mixture of permutations of singlets (e.g. Eq. (\ref{rho3state})),  this allows for a way to obtain the same result but with greatly reduced computational resources.

\subsection{\replaced{Permutation symmetrized Bell state}{Ideal initial state}}

\added{We first show numerical results where the initial state is taken to be the PSB state. }

In Fig. \ref{fig2}(a) and Fig. \ref{fig3}(a) we \deleted{have shown} show the fidelity
\begin{align}
    F = \langle {\cal S}_N | \rho | {\cal S}_N \rangle 
    \label{fidelity}
\end{align}
\replaced{for $ N= 4 $ and $ N = 6 $ respectively, as a function of the}{through various} iterations of our distillation protocol of Sec. \ref{sec:protocol}.  We see that the fidelity quickly approaches the supersinglet state for both \replaced{cases}{$ N = 4,6 $ qubits}.  We also show the success probability (\ref{generalsuccessprob}) which shows that larger systems tend to have a smaller success probability, which is expected as the Hilbert space dimension grows and more outcomes are possible.


\subsection{\replaced{Werner-PSB states}{Noisy initial state}}
\label{sec:noisyinitial}

In Fig. \ref{fig2}(b), we show the effect of starting with a state that is not perfectly in the $ s = 0 $ sector, by starting with an initial Werner state
\begin{align}
    \rho = (1-\epsilon) \rho_{\text{PSB}}  + \epsilon \frac{I}{2^N} .
    \label{wernerstate}
\end{align}
The form of (\ref{wernerstate}) assumes that Steps 1-3 are complete before the depolarizing channel is applied. The mixing with the state $ I/2^N $ results in a population of all spin sectors, including $ s>0 $.  

We see that initially the state approaches the supersinglet, but the fidelity reaches a maximum and then degrades.  The reason for this is that a competing state in the $ s = 1 $ sector starts to develop. Specifically, for $ N = 4$ this state  is $ |1,1,0 \rangle + \sqrt{2} |1,2,0 \rangle \propto |0011\rangle - |1100 \rangle $, which has the same symmetry as the supersinglet.  This state removes the population from the desired supersinglet state.  Unfortunately, twirling (\ref{twirlformula}) cannot remove this state due to the same reason as discussed around Eq. (\ref{twirlingexample}).  This shows the importance of first purifying the state, such as to eliminate undesired spin components using Bell state purification.  In practice, we consider that the parameter $ \epsilon$ can be made small since the preparation Step 1 involves a purification process. 

\added{The peaking behavior occurs for any $ N $, as shown in Fig. \ref{fig3}(d). In Appendix \ref{app:peaking} we show that it is possible to reproduce the peaking behavior with a simple population dynamics model keeping track of the proportions in the supersinglet, and other $ s =0, 1 $ components.  
}

\subsection{\added{Werner-supersinglet states}}

\added{Another possible initial state that can be chosen is the Werner-supersinglet state given by}
\added{
\begin{equation}
\rho_{\mathrm{SS}}
=
(1-\epsilon)\lvert S_N\rangle\langle S_N\rvert
+
\frac{\epsilon}{2^N}I,
\label{supersinglet}
\end{equation}
which is used as the $N$-qubit resource state in the initialization procedure of Sec. \ref{sec:protocol}.  The results are shown in Fig. \ref{fig2}(e) and Fig. \ref{fig3}(e) for $ N=4 $ and $ N=6 $ respectively.  In both cases, we observe an initial increase in the fidelity, followed by a decay. The behavior is similar to the Werner-PSB state, and the peak and decay dynamics has a similar origin.  The Werner-supersinglet state (\ref{supersinglet}) for $ \epsilon> 0 $ adds states with $ s> 0 $, including the $ s= 1 $ sector.  This creates a population of the alternative fixed points which destabilizes the convergence to the supersinglet.  }


\subsection{\added{Purification} Threshold}

\added{In the previous sections we have seen the importance of preparing an initial state in the $ s = 0 $ sector.  In this section, we assume that an incoherent mixture of $ s = 0 $ states can be produced.  The question in this case is how large the initial supersinglet proportion needs to be such that convergence towards the supersinglet can be attained.  }

\deleted{We also investigate the threshold such that convergence towards the supersinglet is obtained. } To investigate this we choose the initial state
\begin{align}
\rho = (1-\delta) \frac{\Pi_0 }{A(N,0)} + \delta | {\cal S}_N \rangle \langle {\cal S}_N  |  .
\label{s0mixture}
\end{align}
where $ \Pi_0 = \sum_{\alpha=1}^{A(N,0)} |0,\alpha,0 \rangle \langle 0,\alpha, 0 | $ is the identity matrix in the $ s = 0 $ sector.  For $ \delta > 0 $, the dominant state in the mixture is the supersinglet state, whereas for  $ \delta < 0 $, the other $ s = 0 $ states dominate. In Fig. \ref{fig2}(c) we show the convergence for various $ \delta $.  When the supersinglet state is the dominant state in the mixture ($\delta >0 $), we see convergence towards the supersinglet state as before.  However, when $\delta <0 $, convergence towards the other singlet states is observed, resulting in a drop of the fidelity.

\begin{figure}[t!]
\includegraphics[width=\linewidth]{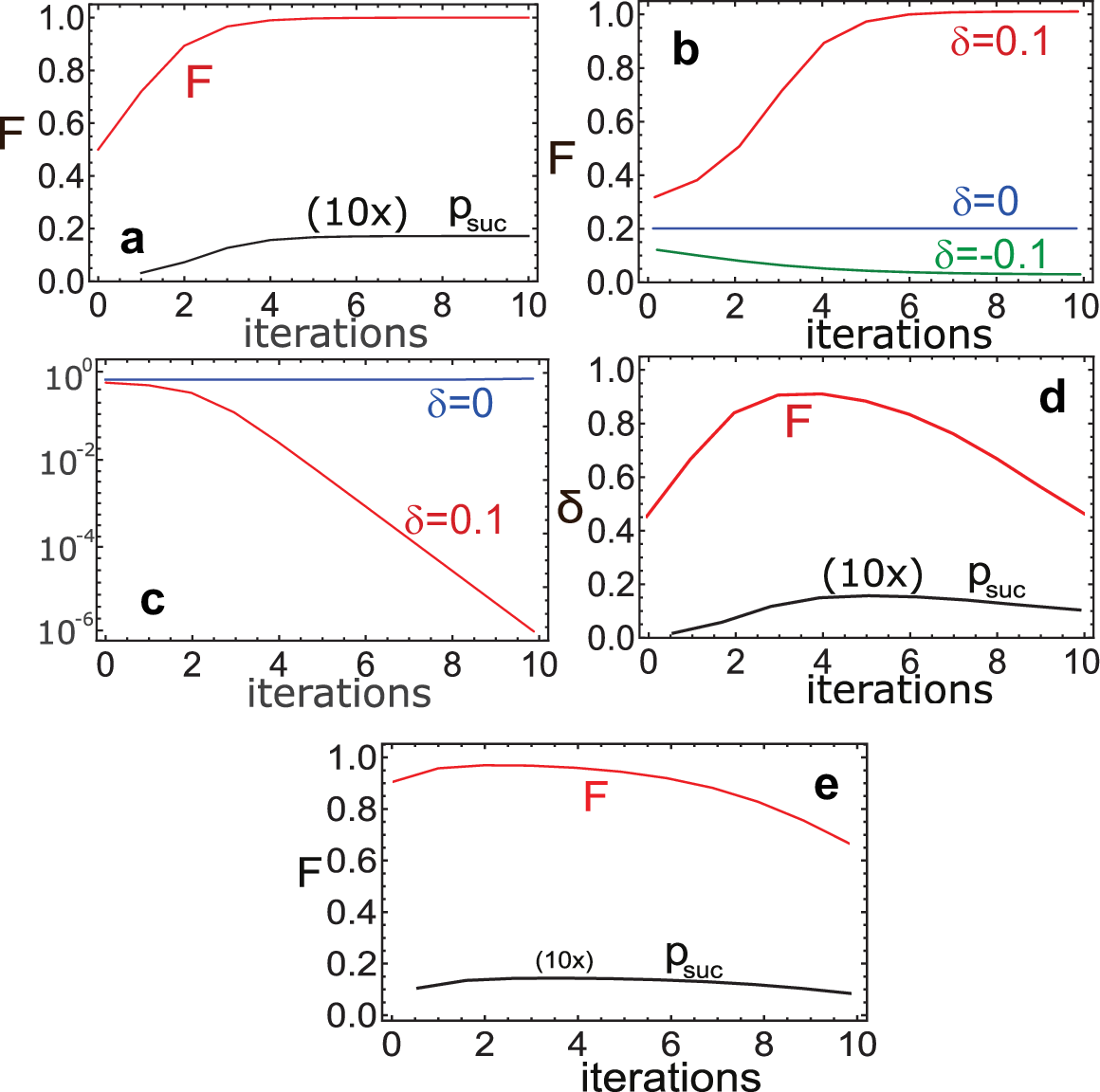}
	\caption{Numerical simulation of \replaced{the}{our} supersinglet distillation protocol \added{of Sec. \ref{sec:protocol} for $ N =6 $}.  \deleted{Each plot shows the fidelity (\ref{fidelity}) and success probability (\ref{generalsuccessprob}) multiplied by a factor of 10 for visibility. (a) $ N = 6 $ with the initial state (\ref{step1state}) with supersinglet symmetries imposed according to (\ref{permsymm}) and (\ref{simpleperm}); (b) fidelities for $ N = 6 $ with the initial state (\ref{s0mixture}) for $ \delta $ as marked; (c) other spin zero states for $ N = 6 $ with the initial state (\ref{s0mixture}) for $ \delta $ as marked on logarithmic scale; (d) fidelities for $ N = 6 $ with the initial state (\ref{wernerstate}) and $ \epsilon = 0.1 $.}
    \added{(a)(d)(e) Fidelity (\ref{fidelity}) and success probability (\ref{generalsuccessprob}) multiplied by a factor of 10 for visibility.  The initial states are chosen as: (a) permutation symmetrized Bell (PSB) state created by starting with (\ref{step1state}), then applying the permutation symmetries according to (\ref{permsymm}) and (\ref{simpleperm}); (b) the spin zero mixture (\ref{s0mixture}) for $ \delta $ as marked; (c) infidelity for the spin zero states (\ref{s0mixture}) for $ \delta $ as marked on logarithmic scale; (d) Werner-PSB state (\ref{wernerstate}) with $ \epsilon = 0.1 $; (e) 
    Werner-supersinglet state (\ref{supersinglet}) with $ \epsilon =0.1 $.  }     
    }
	\label{fig3}%
\end{figure}

In Fig. \ref{fig3}(b) we show results for $ N = 6 $, which shows similar behavior. Again, for $ \delta > 0 $, the fidelity approaches 1, whereas for $ \delta < 0 $ the fidelity collapses to 0.  For both $N = 4$ and 6, $ \delta = 0$ is the threshold, which means there is an exact balance of the convergence and no change in fidelity is seen. We thus conclude that for our supersinglet distillation to converge, it must be the dominant state within the $ s = 0 $ multiplicity.  This  is structurally analogous to BBPSSW Bell purification \cite{bennett1996} where the threshold is $ F =0.5$.  For states of the form (\ref{s0mixture}), the threshold fidelity $ \delta = 0 $ corresponds to $ F = 1/A(N,0) $, which decreases with $ N $.

\subsection{\replaced{GHZ state}{Alternative initial states}}
\label{sec:alternativeinitial}

\added{The PSB state is most likely the simplest choice of initial state that allows for convergence towards the supersinglet state. There are however alternatives which can also be used with the same protocol. } \deleted{The symmetrized singlet state is not the only initial state that works with our distillation protocol. } Another choice is the modified GHZ state
\begin{align}
|\psi_1 \rangle =\left(  \bigotimes_{n=1}^{N/2} X_n \right)  \frac{1}{\sqrt{2}} ( |0 \rangle^{\otimes N} + (-1)^{N/2} |1 \rangle^{\otimes N}  ) .
\label{ghzlikestate}
\end{align}
The above state corresponds to bit flipping the qubits of a GHZ state in group I.  GHZ states are another class of states for which LOCC purification protocols are available \cite{murao1998multiparticle,dur2003multiparticle,kruszynska2006entanglement}.

The state (\ref{ghzlikestate}) is already symmetric with respect to the supersinglet symmetries (\ref{perm1}) and (\ref{perm2}). For this reason, the permutation-symmetrization step, Step 3 of the protocol is unnecessary. However, it is not a pure $s = 0 $ state and twirling operations are required (Step 2). The main feature that makes the state (\ref{ghzlikestate}) suitable is that it lacks components in the odd spin sectors and has a non-zero overlap with the supersinglet state.  Thus the issues with alternative fixed points as discussed in Sec.  (\ref{sec:noisyinitial}) do not occur. More generally, we find that initial states with mixtures of states with even numbers of Pauli bit or phase flips from the supersinglet state converge well under our distillation protocol.  In Fig. \ref{fig2}(d), we show the fidelity and probability evolution with the number of iterations.  We see a similar dependence to the singlet initialization (Fig. \ref{fig2}(a)), with good convergence towards $ F = 1 $.

\section{\added{Experimental realization and scalability}}

The most likely experimental realization is a combination of a local quantum memory at each party's site consisting of 3 qubits, combined with long-distance entanglement distribution realized by photons. The distribution of a Bell pair $ | \Psi^- \rangle $ and its purification is already a widely investigated experimental technique in quantum information, with many realizations
\cite{kwiat2001experimental,pan2001,pan2003experimental,kalb2017entanglement,yin2012}.  After photon distribution, a transducer to transfer the quantum state of the photons to the quantum memory would be also necessary.  Then at each local site, we assume that the operations of Sec. \ref{sec:allowed} can be performed.  The primary operation in our protocol is the measurement following a Schur transformation, which has a complexity scaling as $ O(d^2) $ \cite{bacon2006efficient}.  In our case, the number of copies $ d = 3 $ is fixed, and has been experimentally realized \cite{pivoluska2022implementation}. Apart from this, all operations are local with the exception of the permutations. As discussed in Sec. \ref{sec:enforce}, explicit SWAP operations are not required if the photons are distributed in an appropriate way to begin with.

We expect that our protocol works in principle for any $ N $, but there are bottlenecks when scaling to large $ N $.  While the protocol introduced in Sec. \ref{sec:protocol} only involves LOCC and Bell pair distribution steps, in order to ensure that the correct supersinglet is targeted, some resource intensive steps are required.  Specifically, Step 3 involves performing a permutation symmetrization.  Using the same method as (\ref{rho3state}), this would involve mixing $ (N/2)! $ configurations to impose the correct supersinglet symmetry. \added{We note that the factorial overhead is required only for the state initialization stage and does not need to be repeated with every purification round. Each recurrence step uses only three copies locally stored operating only their local qubits. As a result, the size of the necessary local Schur transformation is independent of $N$.}   We expect the success probability to drop with $ N $, as can be seen by comparing the results for Fig.  \ref{fig2}(a) and Fig. \ref{fig3}(a).  The primary reason for this is that the singlet sector multiplicity increases exponentially with $ N $ according to (\ref{singletmult}). 

\added{While we expect poor scaling with respect to $ N $, this does not necessarily mean that the practical applications are limited.  A key feature of our protocol is that it uses only LOCC.  This means that it is compatible with purification of entangled states that are separated by large distances \cite{yin2017satellite}.  Non-trivial operations are possible even for small $ N $ supersinglets that are distributed over large distances.  For example, in multiparty quantum clock synchronization \cite{oujaa2025phase}, supersinglets are used to synchronize to a central clock.  The advantage of quantum clock synchronization is that the synchronization may proceed without the effects of the intervening medium.  Another application is Byzantine agreement and liar detection \cite{cabello2003solving}, and related protocols have been demonstrated experimentally using four-photon entanglement \cite{gaertner2008experimental}. More recently, the four-qubit singlet has been investigated as a resource for quantum-aided weak broadcast and experimentally prepared on superconducting and trapped-ion quantum-computing platforms \cite{guba2024resource}. }

\section{Summary and Conclusions}
\label{sec:conc}

We have introduced a distillation protocol to purify supersinglet states in qubit systems.  Our main result is the protocol summarized in Sec. \ref{sec:protocol}.  The three copy purification projects onto the total spin basis of the 3 qubits, and we postselect on the outcome corresponding to one of the spin-1/2 irreducible representations.  One of the unusual aspects of our distillation protocol is that it does not start in a noisy version of the supersinglet state, but rather a suitably symmetrized product state of singlet Bell states.  This is highly compatible with existing methods for Bell state distribution \cite{yin2017satellite,neumann2022continuous} and purification \cite{kwiat2001experimental,pan2003experimental,wang2006experimental,yan2022entanglement}.  Rather than a limitation, this is likely to be more convenient than starting from noisy supersinglet states, which would require a more complex algorithm such as the Schur transform \cite{bacon2006efficient} to produce them in the first place. By using conventional Bell state purification, most of the noise introduced during distribution of the state can be removed, leaving a high fidelity initial state that can be input to the recurrence steps of our protocol.  Since standard Bell purification only uses LOCC operations, our whole protocol remains compatible with LOCC. 
One drawback of our approach is the low success probability due to the necessity of postselection at each round.  This results in an exponential decay of the total success probability with the number of rounds of purification. Thus realistically only a few rounds of purification are likely to be possible before the total number of states becomes intractable.  Fortunately, however, as seen in Fig. \ref{fig2}(a) and \ref{fig3}(a), high fidelities are reachable after only several rounds of purification.  Methods to improve the success probability is left as future work.
\added{Despite the increasing resource requirements for large $N$, modestly sized supersinglet states already have experimentally relevant quantum information applications \cite{oujaa2025phase,cabello2003solving,guba2024resource}. The development of our distillation protocol opens the door for applications of the supersinglet state, such as quantum cryptography, quantum clock synchronization, and quantum metrology.} \deleted{The development of our distillation protocol opens the door for applications of the supersinglet state, such as quantum cryptography, quantum clock synchronization, and quantum metrology.}

\begin{acknowledgments}
This work is supported by NYU Shanghai Major-Grants Seed Fund; and Tamkeen under the NYU Abu Dhabi Research Institute grant CG008.
\end{acknowledgments}

\appendix

\section{Local angular momentum basis and angular momentum coupling}
\label{app:3qubit}

The $ j = 3/2, \alpha = 1 $ spin sector eigenstates are 
\begin{align}
| \tfrac{3}{2},1,\tfrac{3}{2} \rangle & = |000 \rangle  \nonumber \\
| \tfrac{3}{2},1,\tfrac{1}{2} \rangle & = \frac{1}{\sqrt{3}} ( |001 \rangle + |010\rangle  + |100 \rangle   ) \nonumber \\
| \tfrac{3}{2},1,-\tfrac{1}{2} \rangle & = \frac{1}{\sqrt{3}} ( |110 \rangle + |101\rangle  + |011 \rangle   ) \nonumber \\
| \tfrac{3}{2},1,-\tfrac{3}{2} \rangle & = |111 \rangle .
\label{spin32}
\end{align}
The first of the $ j = 1/2 $ sector eigenstates with $ \alpha = 1 $ are
\begin{align}
| \tfrac{1}{2},1, \tfrac{1}{2} \rangle & = \frac{1}{\sqrt{6}} (-2 |001 \rangle + |010\rangle + |100 \rangle  )  \nonumber \\
| \tfrac{1}{2},1, -\tfrac{1}{2} \rangle & = \frac{1}{\sqrt{6}} ( - 2|110 \rangle +  | 101\rangle + |011 \rangle ) .
\label{spin121}
\end{align}
This is the postselection basis (\ref{3qubitlogical}) that is used in our supersinglet distillation protocol.  The second $ j = 1/2 $ sector  eigenstates with $ \alpha =2 $  are
\begin{align}
| \tfrac{1}{2},2, \tfrac{1}{2} \rangle & = \frac{1}{\sqrt{2}} (|010 \rangle - | 100\rangle  )\nonumber \\
| \tfrac{1}{2},2, -\tfrac{1}{2} \rangle & = \frac{1}{\sqrt{2}} ( |011 \rangle - | 101\rangle ) .
\label{spin122}
\end{align}

\begin{figure}[t]
\includegraphics[width=\linewidth]{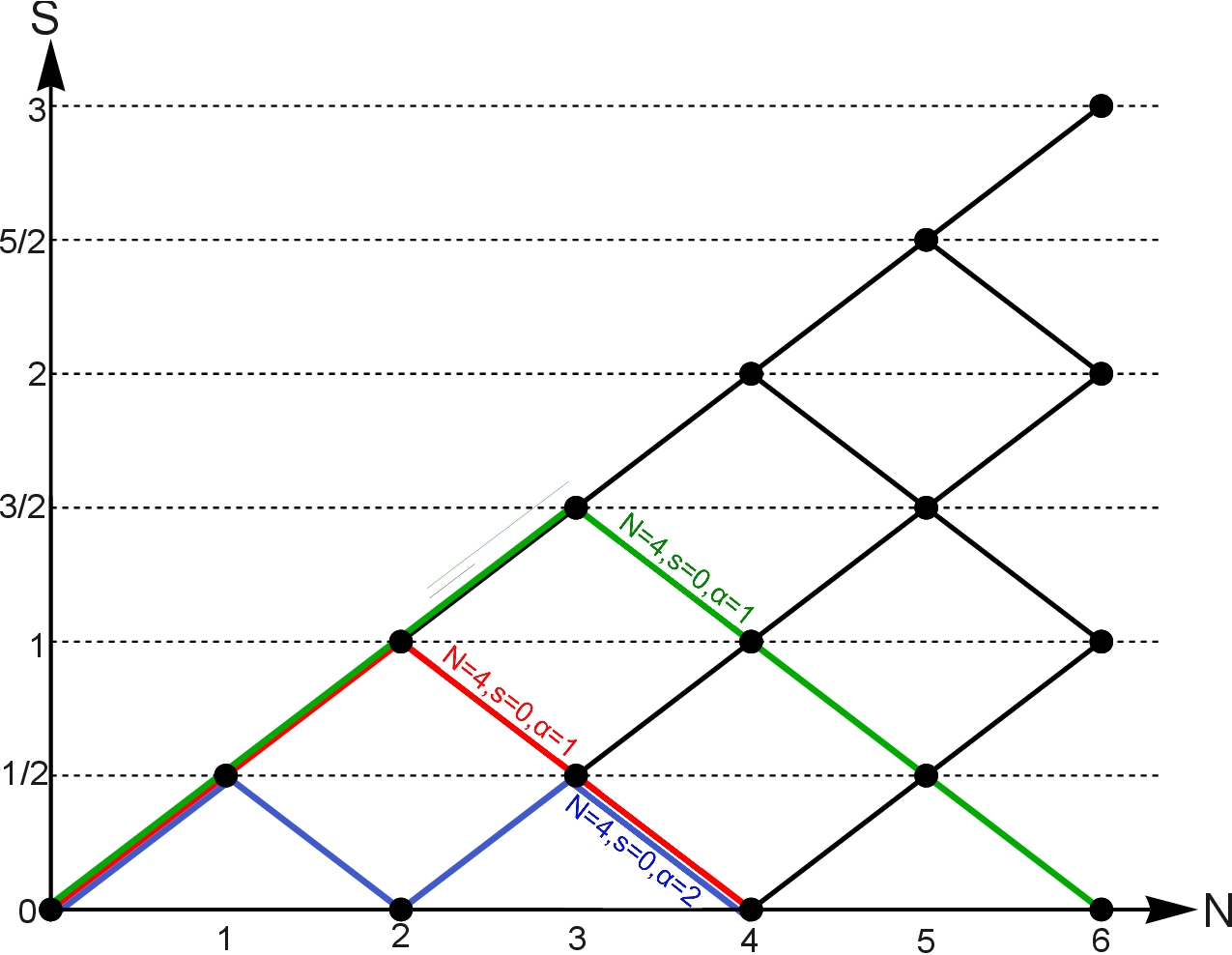}
	\caption{Roadmap of successive angular momentum coupling for spin-$1/2$ particles. Each path represents an irreducible representation of the total angular momentum.}
	\label{fig4}%
\end{figure}

The spin sectors (\ref{spin121}) and (\ref{spin122}) both have total spin $j=1/2$, but they correspond to different outer multiplicity labels because they arise from different angular-momentum coupling paths. Figure \ref{fig4} illustrates this structure. The two paths that end at $N=3$, $j=1/2$ correspond to the two possible multiplicities $\alpha=1,2$.

The same multiplicity structure in Figure \ref{fig4} explains why (\ref{maximalspin}) is needed to identify the supersinglet. For $N=4$, there are two distinct total-spin-zero states. The supersinglet is the one in which the first $N/2$ qubits are coupled symmetrically to the maximal spin $N/4$, the second $N/2$ qubits are also coupled symmetrically to the maximal spin $N/4$, and these two effective spins are then coupled antiferromagnetically to total spin zero. This is precisely the condition stated in (\ref{maximalspin}).

Figure \ref{fig4} also shows the coupling pattern for the $ N = 6 $ supersinglet, which is explicitly given by 
\begin{align}
| {\cal S}_6 \rangle = & \frac{1}{6} ( -3 | 000111 \rangle + |001011\rangle + |001101\rangle + |001110\rangle \nonumber \\ & + |010011\rangle+ |010101\rangle+ |010110\rangle -|011001\rangle \nonumber \\ & -|011010\rangle-|011100\rangle + | 100011 \rangle+ | 100101 \rangle \nonumber \\ & + | 100110 \rangle - | 101001 \rangle- | 101010 \rangle  - | 101100 \rangle \nonumber \\ & - | 110001 \rangle- | 110010\rangle- | 110100 \rangle \nonumber + 3 | 111000\rangle ) .
\end{align}

\section{Schur transform conventions}
\label{vjamordering}


Here we define the computational-basis ordering used in the local Schur transform (\ref{schurunitary}):
\begin{align} 
v(j,\alpha,m) = 5j+2\alpha-m-4 . 
\end{align} 
We label the three-qubit computational basis states by an integer $v\in\{0,\dots,7\}$ with a corresponding binary decomposition. With this convention, the local Schur transform maps the angular-momentum basis to computational-basis labels as
\begin{align} 
| \tfrac{1}{2},1,\tfrac{1}{2}\rangle &\mapsto |0\rangle \equiv |000\rangle, \nonumber \\ 
| \tfrac{1}{2},1,-\tfrac{1}{2}\rangle &\mapsto |1\rangle \equiv |001\rangle, \nonumber \\ 
| \tfrac{1}{2},2,\tfrac{1}{2}\rangle &\mapsto |2\rangle \equiv |010\rangle, \nonumber \\ 
| \tfrac{1}{2},2,-\tfrac{1}{2}\rangle &\mapsto |3\rangle \equiv |011\rangle, \nonumber \\ 
| \tfrac{3}{2},1,\tfrac{3}{2}\rangle &\mapsto |4\rangle \equiv |100\rangle, \nonumber \\ 
| \tfrac{3}{2},1,\tfrac{1}{2}\rangle &\mapsto |5\rangle \equiv |101\rangle, \nonumber \\ 
| \tfrac{3}{2},1,-\tfrac{1}{2}\rangle &\mapsto |6\rangle \equiv |110\rangle, \nonumber \\ 
| \tfrac{3}{2},1,-\tfrac{3}{2}\rangle &\mapsto |7\rangle \equiv |111\rangle . 
\end{align}
The function $v(j,\alpha,m)$ is therefore an indexing convention, specifying which computational basis state is mapped from which angular-momentum basis vector.

\section{Twirling operations}
\label{app:twirl}

Here we prove the relation (\ref{twirlformula}) showing the explicit transformation under twirl operations.  Starting from the definition (\ref{twirldef}), we may write
\begin{align}
T(\rho) = &  \sum_{s l m} \sum_{s' l' m'} \langle s, l, m | \rho | s', l', m '\rangle \nonumber \\
& \times \int dU U^{\otimes N} | s, l, m \rangle \langle s', l', m' | {U^\dagger}^{\otimes N} .
\label{trhostart}
\end{align}
Using the fact that $ U^{\otimes N} = e^{-i \vec{S} \cdot \vec{w} \theta} $, and total spin rotations preserve the $ s $ and $ l $ quantum numbers, we may relate this to the Wigner $D$-matrices \cite{edmonds1996angular}, defined for our purposes as
\begin{align}
D_{\bar{m} m}^{s} (U) = \langle s, l, \bar{m} | U^{\otimes N} | s, l, m \rangle .
\end{align}
Applying this to (\ref{trhostart}), we have
\begin{align}
T(\rho) = &  \sum_{s l m} \sum_{s' l' m'} \langle s, l, m | \rho | s', l', m '\rangle \nonumber \\
& \times \int dU \sum_{\bar{m} \bar{m}'} D_{\bar{m} m}^{s} (U) D_{\bar{m}' m' }^{s'} (U)^*
| s, l, \bar{m} \rangle \langle s', l', \bar{m}' | .
\end{align}
We may now apply the identity \cite{edmonds1996angular}
\begin{align}
 \int dU D_{\bar{m} m}^{s} (U) D_{\bar{m}' m' }^{s'} (U)^* = \frac{\delta_{s s'} \delta_{m m'} \delta_{\bar{m} \bar{m}'} }{2s +1 },
\end{align}
which yields
\begin{align}
T(\rho) = &  \sum_{s l l' } \sum_{m} \frac{\langle s, l, m | \rho | s, l', m \rangle }{2s +1} \sum_{\bar{m}} | s, l, \bar{m} \rangle \langle s, l', \bar{m} | .
\end{align}
With the definition (\ref{gammadef}), we obtain the expression  (\ref{twirlformula}).

\section{Proof that the supersinglet is a fixed point}
\label{app:fixedpointproof}

In this section, we prove Eq. (\ref{fixedpoint}), which states that the supersinglet is a fixed point of the postselected measurement outcome (\ref{psoperatorbig}).  

First, let us write the postselected measurement operator (\ref{psoperatorbig}) as
\begin{align}
{\cal M} = \sum_{\vec{k}} | \vec{k} \rangle 
\langle k_1^{(3)} | \otimes \dots \otimes 
\langle k_N^{(3)} | 
\label{Mopexplicit}
\end{align}
where $ \vec{k} = (k_1, k_2, \dots, k_N) $ and $ k_n \in \{0,1 \} $ specifies a computational basis state, and we used the notation (\ref{3qubitlogical}). Substituting this into (\ref{fixedpoint}) and choosing a particular $  \vec{k} $, we have the equivalent relation
\begin{align}
 \langle \vec{k}^{(3)} | {\cal S}_N \rangle^{\otimes 3}  = \sqrt{p_{\text{suc}}} \langle \vec{k} | {\cal S}_N \rangle
 \label{equivrelation}
\end{align}
where we defined the equivalent computational basis state, but using the 3-qubit $ j = 1/2, \alpha = 1 $ irreducible representation
\begin{align}
| \vec{k}^{(3)} \rangle = |k_1^{(3)} \rangle  \otimes \dots \otimes 
| k_N^{(3)} \rangle .
\end{align}

In Eq. (\ref{measureps}), we defined the three qubit to one qubit map ($ (\mathbb C^2)^{\otimes 3} \to  \mathbb C^2$) that is used in our purification protocol. In fact, it is, an intertwiner since it is a linear map between representation spaces that commutes with SU(2) rotations.  Explicitly, since  $|0^{(3)}\rangle,|1^{(3)}\rangle$ span an irreducible $j=\tfrac12$ subspace that transforms exactly like a physical qubit, one has
\begin{align}
M_{1/2,1} U^{\otimes 3} =  U M_{1/2,1} .
\end{align}
where $U\in\mathrm{SU}(2)$.  
The corresponding $3N$ qubit to $ N $ qubit map for the whole system was defined in (\ref{psoperatorbig}). This is similarly an intertwiner since we have the relation
\begin{align}
{\cal M} U^{\otimes 3N} = U^{\otimes N} {\cal M} .
\label{VtensorNintertwines}
\end{align}

In the $ 3N $ qubit space, $  | {\cal S}_N \rangle^{\otimes 3} $ is a total spin zero state
\begin{align}
S_{3N}^2  | {\cal S}_N \rangle^{\otimes 3} = 0 
\label{totalspin3nzero}
\end{align}
where $ \vec{S}_{3N} =\frac{1}{2} \sum_{n=1}^N \sum_{d=1}^3 \vec{\sigma}_{nd} $,  since it is a product of three spin zero states.  Since ${\cal M}$ is an intertwiner (\ref{VtensorNintertwines}), it maps the $3N$-qubit singlet into the $N$-qubit singlet subspace.  Schur's lemma states that any intertwiner between two irreducible representations of a group is either the zero map or an isomorphism. Both are one-dimensional, so by Schur’s lemma there exists a constant $c\neq 0$ such that
\begin{align}
{\cal M}  | {\cal S}_N \rangle^{\otimes 3} = c |s=0 \rangle .
\label{schurresult}
\end{align}
where $ |s=0\rangle $ is an $N$-qubit spin zero state.  The question now is which spin zero state it is.  

At this point we note that the state $  | {\cal S}_N \rangle^{\otimes 3} $ also has the same permutation symmetries (\ref{perm1}) and (\ref{perm2}) with respect to $ n $ label interchanges
\begin{align}
(P_{\sigma_{\text{I},\text{II}}}  | {\cal S}_N \rangle )^{\otimes 3} & =  | {\cal S}_N \rangle^{\otimes 3} \nonumber \\
(P_{\sigma_{\text{I} \leftrightarrow \text{II} } } | {\cal S}_N \rangle )^{\otimes 3} & = (-1)^{N/2} | {\cal S}_N \rangle^{\otimes 3} .
\label{perm3Nsymm}
\end{align}
With respect to the $ |k_n^{(3)} \rangle $ basis states, the permutation operators $ P_{\sigma_{\text{I},\text{II}}}^{\otimes 3} $ and $ P_{\sigma_{\text{I} \leftrightarrow \text{II} } }^{\otimes 3} $ perform the same interchange as the original $ P_{\sigma_{\text{I},\text{II}}} $ and $ P_{\sigma_{\text{I}\leftrightarrow \text{II}}} $  permutations.  These symmetries select a unique one-dimensional subspace inside the $s=0$ multiplicity sector, which fixes the right hand side of (\ref{schurresult}) to 
\begin{align}
{\cal M}  | {\cal S}_N \rangle^{\otimes 3} = c | {\cal S}_N \rangle.
\label{fixedpoint_with_c}
\end{align}

Taking the overlap of (\ref{fixedpoint_with_c}) with $\langle \vec{k} |$ and using (\ref{Mopexplicit}) we have 
\begin{align}
\langle \vec{k}^{(3)} |{\cal S}_{N}\rangle^{\otimes 3} 
= c \langle \vec{k} |{\cal S}_{N}\rangle .
\end{align}

%
%
\noindent which is precisely the proportionality (\ref{equivrelation}) with a constant independent of $\vec{k}$, and fixes the success probability as
\begin{align}
p_{\mathrm{suc}}= |c|^2 .
\end{align}

\section{Truncation in spin zero sector}
\label{app:simulation} 

In this section we assume that the density matrix $ \rho $ is entirely in the spin zero sector.  Under this assumption, the density matrix satisfies
\begin{align}
\rho = \Pi_0 \rho \Pi_0
\end{align}
where
\begin{align}
\Pi_0 = \sum_{\alpha =1}^{A(N,0)} |\alpha\rangle \langle\alpha|
\end{align}
is the projector in the $ s =0 $ sector and we defined
\begin{align}
  |\alpha \rangle \equiv  | s=0,\alpha,m=0 \rangle 
\end{align}
for notational simplicity. 

In this case, the three copies of the state can be written
\begin{align}
\rho^{\otimes 3} & = ( \Pi_0 \rho \Pi_0)^{\otimes 3} \nonumber \\
& = \sum_{\alpha_1 \alpha_1' \alpha_2 \alpha_2' \alpha_3 \alpha_3'}
\rho_{\alpha_1 \alpha_1'} \rho_{\alpha_2 \alpha_2'} \rho_{\alpha_3 \alpha_3'}
| \alpha_1, \alpha_2,\alpha_3 \rangle \langle \alpha_1' , \alpha_2' ,\alpha_3' | ,
\end{align}
where $ \rho_{\alpha \alpha'} = \langle \alpha | \rho | \alpha' \rangle $.  

Due to the translational invariance of the measurement operator $ \cal M $, the updated density matrix $ \rho' $ in Eq. (\ref{rhoupdate}) is also in the spin-zero sector
\begin{align}
   {\cal M} \rho^{\otimes 3} {\cal M}^\dagger = &
\Pi_0 {\cal M} \rho^{\otimes 3} {\cal M}^\dagger \Pi_0 .
\end{align}
This has matrix elements in the spin zero sector 
\begin{align}
\langle \alpha |  {\cal M} \rho^{\otimes 3} {\cal M}^\dagger  | \alpha' \rangle  =  & \sum_{\vec{k}}   \sum_{\vec{k}' } 
\sum_{\alpha_1 \alpha_1' \alpha_2 \alpha_2' \alpha_3 \alpha_3'}
\rho_{\alpha_1 \alpha_1'} \rho_{\alpha_2 \alpha_2'} \rho_{\alpha_3 \alpha_3'} \nonumber \\
& \times \langle \alpha |  \vec{k} \rangle 
\langle  \vec{k}' | \alpha'  \rangle \Omega_{\vec{k}}^{\alpha_1 \alpha_2 \alpha_3}
 ( \Omega_{\vec{k}'}^{\alpha_1' \alpha_2' \alpha_3'})^*
 \label{updatespinzero}
\end{align}
where we used (\ref{Mopexplicit}) and defined
\begin{align}
 \Omega_{\vec{k}}^{\alpha_1 \alpha_2 \alpha_3} = 
\Big( \langle k_1^{(3)} | \otimes \dots \otimes 
\langle k_N^{(3)} | 
 \Big) | \alpha_1,  \alpha_2,  \alpha_3 \rangle .
\end{align}
Evaluating with the expression (\ref{updatespinzero}) avoids explicitly calculating $ \rho^{\otimes 3} $, reducing the computational resources.

\section{\added{Simple model for peak and decay dynamics of fidelity}}
\label{app:peaking}

\added{The peak value of fidelity can be reproduced by considering a simplified population dynamics model. The model is designed to capture the competition dynamics responsible for the peak and decay behavior seen in the fidelity for initial states with non-zero populations of $ s >0  $ states. We keep track of three populations, the target supersinglet  $N_s(t)$, the spin-1 states $N_1(t)$, and other (non-supersinglet) spin-0 states $N_0(t)$. Our phenomenological model reads}
\added{\begin{align}
\dot{N}_s & = \alpha N_0 - \gamma N_s N_1
\label{spin-s}  \\
\dot{N}_1 & = \beta N_0 N_1 + \gamma N_s N_1
\label{spin-1} \\
\dot{N}_0 & = -\alpha N_0 - \beta N_0 N_1  .
\label{spin-0}
\end{align}}
\added{Here, $\alpha$ is the rate of transfer between $N_{0}$ and $N_{s}$ due to purification, $\beta$ controls the rate of transfer from $N_{0}$ to $N_{1}$, while the term $\gamma$ representing the leakage from target supersinglet $N_{s}$ to $N_{1}$.
The transfers involving $ N_1 $ are nonlinear, in that they are also proportional to $ N_1 $ itself, such that they only occur if there is a non-zero population on $ N_1 $. The evolution equations preserve normalization, which we take as
\begin{equation}
 N_s(n)+N_0(n)+N_1(n)=1 .
  \end{equation}
}

\added{We may solve for $ N_0(t) $ using (\ref{spin-0})} by writing 
\added{\begin{equation}
\int_{N_0(0)}^{N_0(t)} \frac{dN_0}{N_0}
=\int_0^t \left(-\alpha - \beta N_1\right) dt' .
\end{equation}}
\added{Integrating, we have}
\added{\begin{equation}
N_0(t) = N_0(0)\exp\left[
-\alpha t - \beta \int_0^t N_1(u) dt' \right] .
\end{equation}}
\added{Similarly we may solve for (\ref{spin-1}) to give}
\added{
\begin{equation}
N_1(t) = N_1(0)\exp\left[
\int_0^t \left(\beta N_0(t') + \gamma N_s(t')\right) dt'
\right]
\end{equation}}
\added{Integrating (\ref{spin-s}), we have }
\added{\begin{equation}
N_s(t) = N_s(0) +
\int_0^t \left[\alpha N_0(t')-\gamma N_s(t') N_1(t')\right] dt'  .
\end{equation}}
\added{Here, we may interpret $\alpha N_0$ as the gain, and $\gamma N_s N_1 $ as the loss with respect to the supersinglet population.  }

\added{To find the peak of $ N_s(t) $, we wish to find the point where}
\added{\begin{equation}
\frac{dN_s}{dt} = 0.
\end{equation}}
\added{From (\ref{spin-s}), the peak therefore must occur when the gain equals the loss}
\added{\begin{equation}
\alpha N_0(t) = \gamma N_s(t)N_1(t).
\label{gainlosseq}
\end{equation}}

\begin{figure}[t!]
\includegraphics[width=\linewidth]{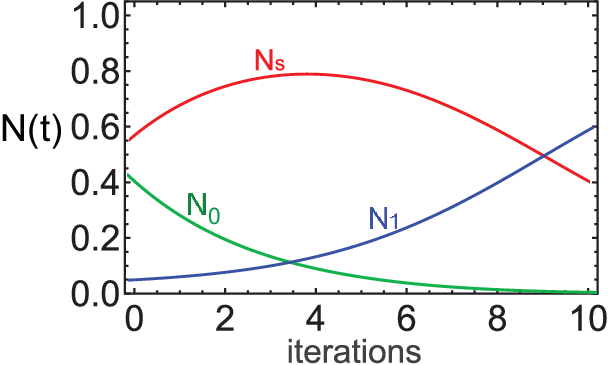}
	\caption{\added{Numerical simulation of the simple population model  (\ref{spin-s})-(\ref{spin-0}). Here the normalized population $N(t)$ is plotted against iterations $t$. The curves $N_s$, $N_0$ and $N_1$ show the populations of the target supersinglet, other spin-0 states and spin-1 states, respectively.  The initial population is taken as $N_s=0.55$, $N_0=0.43$ and $N_1=0.02$, with rate constants chosen as $\alpha=0.30$, $\beta=0.45$ and $\gamma=0.35$. 
    }}
	\label{fig-model}%
\end{figure}

\added{A numerical simulation of population dynamics model is shown in Fig. \ref{fig-model}.  We observe that the population of the target supersinglet $N_s$ increases up to the peak value and then decreases. The spin-0 population $N_0$ depletes exponentially, while the spin-1 population $N_1$ increases in the range of the simulation.  The peak supersinglet population occurs at a value which agrees with (\ref{gainlosseq}).}

\clearpage

\bibliography{ref}

@article{byrnes2012,
    author  = "T. Byrnes and K. Wen and Y. Yamamoto",
    title   = "{Macroscopic quantum computation using Bose-Einstein condensates}",
    year    = "2012",
    journal = "Phys. Rev. A",
    volume  = "85",
    number  = " ",
    pages   = "040306(R)"
}

@article{bennett1996,
	author  = "C. H. Bennett and G. Brassard and S. Popescu and B. Schumacher and J. A. Smolin and W. K. Wootters",
	title   = "Purification of noisy entanglement and faithful teleportation via noisy channel",
	year    = "1996",
	journal = "Phys. Rev. Lett.",
	volume  = "76",
	number  = " ",
	pages   = "722--725"
}

@article{jozsa2000,
	author  = "R. Jozsa and D. S. Abrams and J. P. Dowling and C. P. Williams",
	title   = "Quantum clock synchronisation based on shared prior entanglement",
	year    = "2000",
	journal = "Phys. Rev. Lett.",
	volume  = "85",
	number  = " ",
	pages   = "2010--2013"
}

@article{ilo-okeke2018,
	author  = "E. O. Ilo-Okeke and L. Tessler and J. P. Dowling and T. Byrnes",
	title   = "Remote quantum clock synchronization without synchronized clocks",
	year    = "2018",
	journal = "npj Quantum Inf",
	volume  = "4",
	number  = " ",
	pages   = "40"
}

@article{bennett1996c,
	author  = "C. H. Bennett and H. J. Bernstein and S. Popescu and B. Schumacher",
	title   = "Concentrating partial entanglement by local operations",
	year    = "1996",
	journal = "Phys. Rev. A",
	volume  = "53",
	number  = " ",
	pages   = "2046--2052"
}

@article{lo2001concentrating,
  title={Concentrating entanglement by local actions: Beyond mean values},
  author={Lo, Hoi-Kwong and Popescu, Sandu},
  journal={Physical Review A},
  volume={63},
  number={2},
  pages={022301},
  year={2001},
  publisher={APS}
}

@article{pan2001,
	author  = "J. W. Pan and C. Simon and C. Brukner and A. Zeilinger",
	title   = "Entanglement purification for quantum communication",
	year    = "2001",
	journal = "Nature",
	volume  = "410",
	number  = " ",
	pages   = "1067--1070"
}

@article{yin2012,
	author  = "J. Yin and J. G. Ren and H.  Lu and Y. Cao and H. L. Yong and Y. P. Wu and C.  Liu and S. K. Liao and F. Zhou and Y.  Jiang and X. D. Cai and P.  Xu and G. S. Pan and J. J.  Jia and Y. M. Huang and H.  Yin and J. Y.  Wang and Y. A. Chen and C. Z.  Peng and J. W. Pan",
	title   = "{Quantum teleportation and entanglement distribution over 100-kilometre free-space channels}",
	year    = "2012",
	journal = "Nature ",  
	volume  = "488",
	number  = " ",
	pages   = "185--188"
}

@article{maneva2000,
	author  = "E. N. Maneva and J. A. Smolin",
	title   = "Improved two-party and multi-party purification protocols",
	year    = "2000",
	journal = "arXiv",
	volume  = "quant-ph",
	number  = " ",
	pages   = "0003099v1"
}

@article{cabello2003supersinglets,
  title={Supersinglets},
  author={Cabello, Ad{\'a}n},
  journal={Journal of Modern Optics},
  volume={50},
  number={6-7},
  pages={1049--1061},
  year={2003},
  publisher={Taylor \& Francis}
}

@article{toth2010generation,
  title={Generation of macroscopic singlet states in atomic ensembles},
  author={T{\'o}th, G{\'e}za and Mitchell, Morgan W},
  journal={New Journal of Physics},
  volume={12},
  number={5},
  pages={053007},
  year={2010},
  publisher={IOP Publishing}
}

@article{pyrkov2014full,
  title={Full-Bloch-sphere teleportation of spinor Bose-Einstein condensates and spin ensembles},
  author={Pyrkov, Alexey N and Byrnes, Tim},
  journal={Physical Review A},
  volume={90},
  number={6},
  pages={062336},
  year={2014},
  publisher={APS}
}

@article{bennett1996purification,
  title={Purification of noisy entanglement and faithful teleportation via noisy channels},
  author={Bennett, Charles H and Brassard, Gilles and Popescu, Sandu and Schumacher, Benjamin and Smolin, John A and Wootters, William K},
  journal={Physical review letters},
  volume={76},
  number={5},
  pages={722},
  year={1996},
  publisher={APS}
}

@article{horodecki1999general,
  title={General teleportation channel, singlet fraction, and quasidistillation},
  author={Horodecki, Micha{\l} and Horodecki, Pawe{\l} and Horodecki, Ryszard},
  journal={Physical Review A},
  volume={60},
  number={3},
  pages={1888},
  year={1999},
  publisher={APS}
}

@article{abdelrahman2014coherent,
  title={Coherent all-optical control of ultracold atoms arrays in permanent magnetic traps},
  author={Abdelrahman, Ahmed and Mukai, Tetsuya and H{\"a}ffner, Hartmut and Byrnes, Tim},
  journal={Optics express},
  volume={22},
  number={3},
  pages={3501--3513},
  year={2014},
  publisher={Optical Society of America}
}

@article{cabello2003solving,
  title={Solving the liar detection problem using the four-qubit singlet state},
  author={Cabello, Ad{\'a}n},
  journal={Physical Review A},
  volume={68},
  number={1},
  pages={012304},
  year={2003},
  publisher={APS}
}

@article{PhysRevA.106.033314,
  title = {Deterministic preparation of supersinglets with collective spin projections},
  author = {Ilo-Okeke, Ebubechukwu O. and Ji, Yangxu and Chen, Ping and Mao, Yuping and Kondappan, Manikandan and Ivannikov, Valentin and Xiao, Yanhong and Byrnes, Tim},
  journal = {Phys. Rev. A},
  volume = {106},
  issue = {3},
  pages = {033314},
  numpages = {12},
  year = {2022},
  month = {Sep},
  publisher = {American Physical Society},
  doi = {10.1103/PhysRevA.106.033314},
  url = {https://link.aps.org/doi/10.1103/PhysRevA.106.033314}
}

@article{chaudhury2007quantum,
  title={Quantum control of the hyperfine spin of a Cs atom ensemble},
  author={Chaudhury, Souma and Merkel, Seth and Herr, Tobias and Silberfarb, Andrew and Deutsch, Ivan H and Jessen, Poul S},
  journal={Physical Review Letters},
  volume={99},
  number={16},
  pages={163002},
  year={2007},
  publisher={APS}
}

@article{deutsch1996quantum,
  title={Quantum privacy amplification and the security of quantum cryptography over noisy channels},
  author={Deutsch, David and Ekert, Artur and Jozsa, Richard and Macchiavello, Chiara and Popescu, Sandu and Sanpera, Anna},
  journal={Physical review letters},
  volume={77},
  number={13},
  pages={2818},
  year={1996},
  publisher={APS}
}

@article{murao1998multiparticle,
  title={Multiparticle entanglement purification protocols},
  author={Murao, M and Plenio, MB and Popescu, Sandu and Vedral, V and Knight, PL},
  journal={Physical Review A},
  volume={57},
  number={6},
  pages={R4075},
  year={1998},
  publisher={APS}
}

@article{miyake2005distillation,
  title={Distillation of multipartite entanglement by complementary stabilizer measurements},
  author={Miyake, Akimasa and Briegel, Hans J},
  journal={Physical review letters},
  volume={95},
  number={22},
  pages={220501},
  year={2005},
  publisher={APS}
}

@article{childs2025streaming,
  title={Streaming quantum state purification},
  author={Childs, Andrew M and Fu, Honghao and Leung, Debbie and Li, Zhi and Ozols, Maris and Vyas, Vedang},
  journal={Quantum},
  volume={9},
  pages={1603},
  year={2025},
  publisher={Verein zur F{\"o}rderung des Open Access Publizierens in den Quantenwissenschaften}
}

@article{li2024optimal,
  title={Optimal quantum purity amplification},
  author={Li, Zhaoyi and Fu, Honghao and Isogawa, Takuya and Chuang, Isaac},
  journal={arXiv preprint arXiv:2409.18167},
  year={2024}
}

@article{bacon2006efficient,
  title={Efficient quantum circuits for Schur and Clebsch-Gordan transforms},
  author={Bacon, Dave and Chuang, Isaac L and Harrow, Aram W},
  journal={Physical review letters},
  volume={97},
  number={17},
  pages={170502},
  year={2006},
  publisher={APS}
}

@article{kwiat2001experimental,
  title={Experimental entanglement distillation and ‘hidden’non-locality},
  author={Kwiat, Paul G and Barraza-Lopez, Salvador and Stefanov, Andre and Gisin, Nicolas},
  journal={Nature},
  volume={409},
  number={6823},
  pages={1014--1017},
  year={2001},
  publisher={Nature Publishing Group UK London}
}

@article{pan2003experimental,
  title={Experimental entanglement purification of arbitrary unknown states},
  author={Pan, Jian-Wei and Gasparoni, Sara and Ursin, Rupert and Weihs, Gregor and Zeilinger, Anton},
  journal={Nature},
  volume={423},
  number={6938},
  pages={417--422},
  year={2003},
  publisher={Nature Publishing Group UK London}
}

@article{wang2006experimental,
  title={Experimental entanglement distillation of two-qubit mixed states under local operations},
  author={Wang, Zhi-Wei and Zhou, Xiang-Fa and Huang, Yun-Feng and Zhang, Yong-Sheng and Ren, Xi-Feng and Guo, Guang-Can},
  journal={Physical review letters},
  volume={96},
  number={22},
  pages={220505},
  year={2006},
  publisher={APS}
}

@article{yan2022entanglement,
  title={Entanglement purification and protection in a superconducting quantum network},
  author={Yan, Haoxiong and Zhong, Youpeng and Chang, Hung-Shen and Bienfait, Audrey and Chou, Ming-Han and Conner, Christopher R and Dumur, {\'E}tienne and Grebel, Joel and Povey, Rhys G and Cleland, Andrew N},
  journal={Physical Review Letters},
  volume={128},
  number={8},
  pages={080504},
  year={2022},
  publisher={APS}
}

@article{yin2017satellite,
  title={Satellite-based entanglement distribution over 1200 kilometers},
  author={Yin, Juan and Cao, Yuan and Li, Yu-Huai and Liao, Sheng-Kai and Zhang, Liang and Ren, Ji-Gang and Cai, Wen-Qi and Liu, Wei-Yue and Li, Bo and Dai, Hui and others},
  journal={Science},
  volume={356},
  number={6343},
  pages={1140--1144},
  year={2017},
  publisher={American Association for the Advancement of Science}
}

@article{neumann2022continuous,
  title={Continuous entanglement distribution over a transnational 248 km fiber link},
  author={Neumann, Sebastian Philipp and Buchner, Alexander and Bulla, Lukas and Bohmann, Martin and Ursin, Rupert},
  journal={Nature Communications},
  volume={13},
  number={1},
  pages={6134},
  year={2022},
  publisher={Nature Publishing Group UK London}
}

@article{browne2003driving,
  title={Driving non-Gaussian to Gaussian states with linear optics},
  author={Browne, Daniel E and Eisert, Jens and Scheel, Stefan and Plenio, Martin B},
  journal={Physical Review A},
  volume={67},
  number={6},
  pages={062320},
  year={2003},
  publisher={APS}
}

@article{eisert2004distillation,
  title={Distillation of continuous-variable entanglement with optical means},
  author={Eisert, Jens and Browne, Dan E and Scheel, Stefan and Plenio, Martin B},
  journal={Annals of Physics},
  volume={311},
  number={2},
  pages={431--458},
  year={2004},
  publisher={Elsevier}
}

@article{bravyi2005universal,
  title={Universal quantum computation with ideal Clifford gates and noisy ancillas},
  author={Bravyi, Sergey and Kitaev, Alexei},
  journal={Physical Review A—Atomic, Molecular, and Optical Physics},
  volume={71},
  number={2},
  pages={022316},
  year={2005},
  publisher={APS}
}

@book{edmonds1996angular,
  title={Angular momentum in quantum mechanics},
  author={Edmonds, Alan Robert},
  volume={4},
  year={1996},
  publisher={Princeton university press}
}

@article{bennett1996mixed,
  title={Mixed-state entanglement and quantum error correction},
  author={Bennett, Charles H and DiVincenzo, David P and Smolin, John A and Wootters, William K},
  journal={Physical Review A},
  volume={54},
  number={5},
  pages={3824},
  year={1996},
  publisher={APS}
}

@article{schlingemann2001quantum,
  title={Quantum error-correcting codes associated with graphs},
  author={Schlingemann, Dirk and Werner, Reinhard F},
  journal={Physical Review A},
  volume={65},
  number={1},
  pages={012308},
  year={2001},
  publisher={APS}
}

@article{dur2003multiparticle,
  title={Multiparticle entanglement purification for graph states},
  author={D{\"u}r, Wolfgang and Aschauer, Hans and Briegel, H-J},
  journal={Physical review letters},
  volume={91},
  number={10},
  pages={107903},
  year={2003},
  publisher={APS}
}

@article{kruszynska2006entanglement,
  title={Entanglement purification protocols for all graph states},
  author={Kruszynska, Caroline and Miyake, Akimasa and Briegel, Hans J and D{\"u}r, Wolfgang},
  journal={Physical Review A—Atomic, Molecular, and Optical Physics},
  volume={74},
  number={5},
  pages={052316},
  year={2006},
  publisher={APS}
}

@article{matsumoto2003conversion,
  title={Conversion of a general quantum stabilizer code to an entanglement distillation protocol},
  author={Matsumoto, Ryutaroh},
  journal={Journal of Physics A: Mathematical and General},
  volume={36},
  number={29},
  pages={8113},
  year={2003},
  publisher={IOP Publishing}
}

@article{aschauer2005multiparticle,
  title={Multiparticle entanglement purification for two-colorable graph states},
  author={Aschauer, Hans and D{\"u}r, Wolfgang and Briegel, H-J},
  journal={Physical Review A—Atomic, Molecular, and Optical Physics},
  volume={71},
  number={1},
  pages={012319},
  year={2005},
  publisher={APS}
}

@article{glancy2006entanglement,
  title={Entanglement purification of any stabilizer state},
  author={Glancy, Scott and Knill, Emanuel and Vasconcelos, Hilma M},
  journal={Physical Review A—Atomic, Molecular, and Optical Physics},
  volume={74},
  number={3},
  pages={032319},
  year={2006},
  publisher={APS}
}

@article{kalb2017entanglement,
  title={Entanglement distillation between solid-state quantum network nodes},
  author={Kalb, Norbert and Reiserer, Andreas A and Humphreys, Peter C and Bakermans, Jacob JW and Kamerling, Sten J and Nickerson, Naomi H and Benjamin, Simon C and Twitchen, Daniel J and Markham, Matthew and Hanson, Ronald},
  journal={Science},
  volume={356},
  number={6341},
  pages={928--932},
  year={2017},
  publisher={American Association for the Advancement of Science}
}

@article{horodecki1998mixed,
  title={Mixed-state entanglement and distillation: Is there a “bound” entanglement in nature?},
  author={Horodecki, Micha{\l} and Horodecki, Pawe{\l} and Horodecki, Ryszard},
  journal={Physical Review Letters},
  volume={80},
  number={24},
  pages={5239},
  year={1998},
  publisher={APS}
}

@article{alber2001efficient,
  title={Efficient bipartite quantum state purification in arbitrary dimensional Hilbert spaces},
  author={Alber, Gernot and Delgado, Aldo and Gisin, Nicolas and Jex, Igor},
  journal={Journal of Physics A: Mathematical and General},
  volume={34},
  number={42},
  pages={8821},
  year={2001},
  publisher={IOP Publishing}
}

@article{vollbrecht2003efficient,
  title={Efficient distillation beyond qubits},
  author={Vollbrecht, Karl Gerd H and Wolf, Michael M},
  journal={Physical Review A},
  volume={67},
  number={1},
  pages={012303},
  year={2003},
  publisher={APS}
}

@article{pivoluska2022implementation,
  title={Implementation of quantum compression on IBM quantum computers},
  author={Pivoluska, Matej and Plesch, Martin},
  journal={Scientific Reports},
  volume={12},
  number={1},
  pages={5841},
  year={2022},
  publisher={Nature Publishing Group UK London}
}

@article{gaertner2008experimental,
  title={Experimental demonstration of a quantum protocol for byzantine agreement and liar detection},
  author={Gaertner, Sascha and Bourennane, Mohamed and Kurtsiefer, Christian and Cabello, Ad{\'a}n and Weinfurter, Harald},
  journal={Physical review letters},
  volume={100},
  number={7},
  pages={070504},
  year={2008},
  publisher={APS}
}

@article{guba2024resource,
  title={Resource analysis for quantum-aided Byzantine agreement with the four-qubit singlet state},
  author={Guba, Zolt{\'a}n and Finta, Istv{\'a}n and Budai, {\'A}kos and Farkas, L{\'o}r{\'a}nt and Zimbor{\'a}s, Zolt{\'a}n and P{\'a}lyi, Andr{\'a}s},
  journal={Quantum},
  volume={8},
  pages={1324},
  year={2024},
  publisher={Verein zur F{\"o}rderung des Open Access Publizierens in den Quantenwissenschaften}
}

@article{oujaa2025phase,
  title={Phase assumption-free multiparty quantum clock synchronization},
  author={Oujaa, Hatim A and Liu, Qiao and Ilo-Okeke, Ebubechukwu O and Ivannikov, Valentin and Dowling, Jonathan P and Byrnes, Tim},
  journal={arXiv preprint arXiv:2510.16895},
  year={2025}
}

\end{document}